\documentclass[preprint2]{aastex}

\newcommand{\myemail}{trawle@as.arizona.edu}

\shorttitle{FIR-Detected Star Formation in Brightest Cluster Galaxies}
\shortauthors{Rawle et al.}

\begin{document}

\title{The Relation Between Cool Cluster Cores and Herschel-Detected Star Formation in Brightest Cluster Galaxies\altaffilmark{}}

\author{T.~D.~Rawle\altaffilmark{1}, A.~C.~Edge\altaffilmark{2}, E.~Egami\altaffilmark{1}, M.~Rex\altaffilmark{1}, G.~P.~Smith\altaffilmark{3}, B.~Altieri\altaffilmark{4}, A.~Fiedler\altaffilmark{1}, C.~P.~Haines\altaffilmark{1,3}, M.~J.~Pereira\altaffilmark{1}, P.~G.~P\'{e}rez-Gonz\'{a}lez\altaffilmark{5,6}, J.~Portouw\altaffilmark{1}, I.~Valtchanov\altaffilmark{4}, G.~Walth\altaffilmark{1}, P.~P.~van~der~Werf\altaffilmark{7}, M.~Zemcov\altaffilmark{8,9}}

\email{\myemail}

\altaffiltext{}{Herschel is an ESA space observatory with science instruments provided by European-led Principal Investigator consortia and with important participation from NASA.}

\altaffiltext{1}{Steward Observatory, University of Arizona, 933 N. Cherry Ave, Tucson, AZ 85721}
\altaffiltext{2}{Institute for Computational Cosmology, Durham University, South Road, Durham, DH1 3LE, UK}
\altaffiltext{3}{School of Physics and Astronomy, University of Birmingham, Edgbaston, Birmingham, B15 2TT, UK}
\altaffiltext{4}{Herschel Science Centre, ESAC, ESA, PO Box 78, Villanueva de la Ca\~{n}ada, 28691 Madrid, Spain}
\altaffiltext{5}{Departamento de Astrof\'{\i}sica, Facultad de CC. F\'{\i}sicas,Universidad Complutense de Madrid, E-28040 Madrid, Spain}
\altaffiltext{6}{Associate Astronomer at Steward Observatory, University of Arizona}
\altaffiltext{7}{Sterrewacht Leiden, Leiden University, PO Box 9513, 2300 RA, Leiden, the Netherlands}
\altaffiltext{8}{Department of Physics, Mathematics and Astronomy,
 California Institute of Technology, Pasadena, CA 91125, USA}
\altaffiltext{9}{Jet Propulsion Laboratory (JPL), National Aeronautics
 and Space Administration (NASA), Pasadena, CA 91109, USA}

%%%%%%%%%%%%%%%%%%%

\begin{abstract}
{We present far-infrared (FIR) analysis of 68 Brightest Cluster Galaxies (BCGs) at 0.08 $<$ $z$ $<$ 1.0. Deriving total infrared luminosities directly from {\it Spitzer} and Herschel photometry spanning the peak of the dust component (24--500 \micron), we calculate the obscured star formation rate (SFR). 22$^{+6.2}_{-5.3}$\% of the BCGs are detected in the far-infrared, with SFR $=$ 1--150 M$_\sun$ yr$^{-1}$. The infrared luminosity is highly correlated with cluster X-ray gas cooling times for cool-core clusters (gas cooling time $<$1 Gyr), strongly suggesting that the star formation in these BCGs is influenced by the cluster-scale cooling process. The occurrence of the molecular gas tracing H$\alpha$ emission is also correlated with obscured star formation. For all but the most luminous BCGs ($L_{\rm TIR}$ $>$ 2$\times$10$^{11}$ $L_{\odot}$), only a small ($\la$0.4 mag) reddening correction is required for SFR(H$\alpha$) to agree with SFR$_{\rm FIR}$. The relatively low H$\alpha$ extinction (dust obscuration), compared to values reported for the general star-forming population, lends further weight to an alternate (external) origin for the cold gas. Finally, we use a stacking analysis of non-cool-core clusters to show that the majority of the fuel for star formation in the FIR-bright BCGs is unlikely to originate form normal stellar mass loss.}
\end{abstract}

\keywords{Galaxies: clusters: general -- Galaxies: elliptical and lenticular, cD -- galaxies: star formation -- infrared: galaxies}

\section{Introduction}

In the central regions of rich galaxy clusters, the intracluster medium can be sufficiently dense that cooling to 10$^4$ K occurs on timescales shorter than the cluster lifetime \citep{cow77-723,fab77-479,edg92-177}. However, X-ray observations indicate that at low gas temperatures (a factor of a few below ambient), the cooling rates are slower than expected for a simple run-away cooling flow model \citep[e.g.][]{tam01-87,pet06-1}. Active Galactic Nuclei (AGN) outbursts could inject sufficient energy into the intracluster medium to suppress cooling. Although the underlying physics is still not well understood, observation of radio bubbles, so called as they are regions relatively empty of thermal X-ray gas coincident with the radio lobes \citep[e.g.][]{boh93-25,mcn00-135}, support AGN heating as a viable mechanism \citep{edg92-177,mcn07-117}.

Many clusters possess a large elliptical galaxy at the minimum of the potential well. These Brightest Cluster Galaxies (BCGs) are often more than a magnitude brighter in optical bands than the second ranked galaxy. In contrast to the majority of massive cluster ellipticals, which are optically red and quiescent, some BCGs contain significant cool gas and exhibit signs of star formation. These BCGs exhibit far-infrared atomic cooling lines \citep{mit11-p}, and millimeter molecular transitions (H$_2$ and CO emission) that reveal large quantities of gas at a range of cool temperatures (T $\la$ 2000 K; \citealt{edg02-49,ega06-21,joh07-1246,edg10-47}). Strong optical emission lines are often used as a proxy for this molecular gas \citep[e.g.][]{cav08-107}, with measured line ratios typical of HII regions \citep{cra99-857,con01-2281}. H$\alpha$ imaging of nearby BCGs  appears to show emission tracing filamentary structure well beyond the central AGN, but apparently created by the radio bubbles \citep[e.g.][]{cra05-216}.

The infrared dust component offers an independent probe of the star formation rate (SFR). In the {\it Spitzer} era, great sensitivity at 24 \micron{} offered the ability to estimate SFR for large samples of BCGs \citep[e.g.][]{ega06-922,qui08-39}. For instance, using the latter of these samples, \citet{ode08-1035} found that the cluster X-ray emission is correlated with IR luminosity for H$\alpha$-detected BCGs, suggesting that cool gas originally associated with the intracluster medium is fueling star formation in BCGs. This interplay is still not well understood, with only a small fraction of the available cold gas forming stars ($\la$ 100 M$_{\odot}$yr$^{-1}$).

With the advent of sensitive far-infrared (FIR) observations from the Herschel Space Observatory, it is now, for the first time, possible for a direct measurement of the dust component in a large sample of BCGs. This study uses Herschel photometry to calculate the total IR luminosity for BCGs in clusters at $z$ $\sim$ 0.08--1.0. Dust properties and SFR can be constrained tightly, without the need to estimate from the mid-infrared. This is particularly important at these redshifts, as the varying intensity of PAH and silicate absorption features within the 24 \micron{} band causes non-systemmatic uncertainty in the extrapolated IR luminosity.

The paper is organized as follows. Section \ref{sec:obs} details the far-infrared observations, while Section \ref{sec:phot} describes the important process of matching flux at different wavelengths (counterparting), and tabulates the final photometry. Section \ref{sec:results} describes the SED template fitting procedure and presents the integrated IR luminosities.  In Section \ref{sec:disc} we discuss the dust and star formation properties in the context of cool-core cluster indicators. The primary conclusions are summarized in Section \ref{sec:sum}.

\section{Observations}
\label{sec:obs}

This study employs far-infrared data from the ESA Herschel Space Observatory \citep{pil10-1}. Specifically, we use photometry from the Photodetector Array Camera and Spectrometer (PACS; \citealt{pog10-2})  at 70, 100, 160 \micron, and from the Spectral and Photometric Imaging Receiver (SPIRE; \citealt{gri10-3}) at 250, 350, 500 \micron. These bands straddle the peak of the far-infrared spectral energy distribution (SED), for sources up to z$\sim$3. The respective beam sizes are larger with increasing wavelength, with FWHM $=$ 5.2", 7.7", 12", 18", 25", 36".

Three Herschel key programs contribute to the full sample of 67 cluster fields (68 BCGs): Herschel Lensing Survey (HLS; \citealt{ega10-12}), Local Cluster Substructure Survey (LoCuSS; \citealt{smi10-18}) and a study of BCGs in known cool-core clusters \citep[][E10]{edg10-47}. Herschel data were processed using an augmented version of the standard reduction pipelines (HIPE; \citealt{ott10-139}). Major modifications are as follows. During PACS reduction, high-pass filtering of the time-stream is necessary to remove the 1/f noise drift. In an attempt to avoid bright source ringing, we implement a simple masking algorithm prior to filtering. In all Herschel processing, signal-to-noise was maximized by including data frames beyond the nominal science scan legs, based on instantaneous scan speed: within 10\% of science scan speed for PACS; $\ga$0.5"/s for SPIRE. In reality, turnaround frames only increase sensitivity at the BCG position for HLS PACS, where the total map size is comparable to a single array footprint. For further details on the Herschel data reduction, please refer to the individual survey papers.

The infrared SEDs for each BCG also include, where available, data from the \textit{Spitzer} archive (IRAC and MIPS), the 2MASS 2.2 \micron{} $K$-band catalog \citep{skr06-1163} and the WISE \citep{wri10-1868} preliminary data release (at 3.4, 4.6, 12, and 22 \micron). Although coverage in these near/mid-infrared bands is not complete, none of our conclusions are sensitive to their availability.

\subsection{Herschel Lensing Survey (HLS)}

HLS \citep{ega10-12} was primarily devised to exploit the lensing effect of massive clusters, probing beyond the confusion limit of the Herschel instruments to observe intrinsically faint, high-redshift sources \citep[e.g.][]{rex10-13}. HLS comprises deep Herschel data in 5 bands, 100--500 \micron. PACS imaging at 100 and 160 \micron{} has mean 3$\,\sigma$ limits of 2.4 and 4.7 mJy, while in the three SPIRE bands, the typical 3$\,\sigma$ limits are 9.4, 10.6, 12.0 mJy respectively.

The survey observed 44 clusters (0.2 $<$ $z$ $<$ 1.0, with the majority at 0.2 $<$ $z$ $<$ 0.4) centered on the diffuse X-ray peak. More than three quarters of the clusters are covered by \textit{Spitzer} MIPS 24 \micron{} imaging, nearly all have IRAC data at 3.6 and 4.5 \micron, while approximately a third also have 5.8, and 8 \micron. Crucially for BCG identification, all of the cluster cores have been observed in at least one HST band.

\subsection{Local Cluster Substructure Survey (LoCuSS)}

LoCuSS is a multi-wavelength (X-ray, through far-UV, optical, far-IR, to radio), wide-field survey of X-ray luminous galaxy clusters at 0.15 $<$ z $<$ 0.3. One of the main motivations is to probe the physics responsible for the transformation of galaxies from late-- to early--types in the cluster in-fall regions and surrounding large-scale structure \citep{smi10-18}.  A key dataset supporting this science goal is Herschel observations at 100--500 \micron{} of a representative sample of 31 clusters. The data covers a wider field of view ($\sim$25'$\times$25') than HLS, to mean 3 $\sigma$ depths of 13.0, 17.0, 14.0, 18.9, 20.4 mJy (100--500 \micron).

The 31 cluster Herschel sample is referred to in this paper as LoCuSS. Eleven of the clusters are also in HLS, and for these, we adopt the deeper HLS Herschel data. The LoCuSS sample has full coverage in the optical from both HST and Subaru Suprime-Cam \citep{oka10-811}. \textit{Spitzer} MIPS 24 \micron{} photometry is complete (\citealt{hai09-126}, Haines et al. in prep.), while IRAC imaging is available for more than half of the clusters. MIPS 70 \micron{} photometry was published for three LoCuSS BCGs in \citet{qui08-39}.

\subsection{Known cool-core clusters (E10)}

The third component explores far-infrared line emission and continuum properties for a sample of 11 well-studied BCGs in cool-core clusters. The BCGs were selected by optical emission and X-ray properties as the most likely FIR-luminous BCGs in the local Universe ($z$ $\la$ 0.3). The Herschel observations pertinent for this study were published in \citet[][E10]{edg10-47} and includes all 6 PACS+SPIRE bands (70--500 \micron) for three BCGs (A1068, Z3146, A2597). These sources have extensive existing optical and infrared photometry, as well as submillimeter observations from JCMT/SCUBA.

\section{Multi-band photometry}
\label{sec:phot}

In this section, we describe the process of allocating multi-band infrared flux to each optical BCG. This applies to the 64 clusters in the HLS and LoCuSS samples. Photometry for the remaining three BCGs was taken directly from E10.

\subsection{BCG identification}

For each cluster, the first task was to identify the BCG, which we define as the optically brightest source in the cluster. We verified literature positions by examining all available optical data, including central HST imaging in every case. Two of the clusters are generally considered to consist of two components, each with their own BCG: A773 \citep{cra99-857} and A1758 \citep{dav04-831}. We confirm this selection, and retain both sources for analysis (labelled north and south). Our conclusions are independent of the inclusion of any combination of these BCGs. MACS0025.4--1222 (the ``Baby Bullet'') lacks an obvious BCG, and is removed from the sample. The left-most columns in Table \ref{tab:phot} present the clusters and supply optical positions for the 65 BCGs (including two pairs).

\begin{figure*}
\includegraphics[angle=0,scale=0.37]{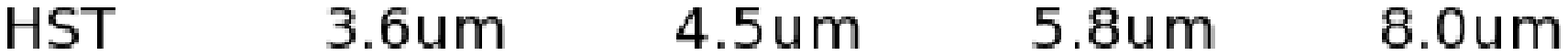}
\includegraphics[angle=0,scale=0.37]{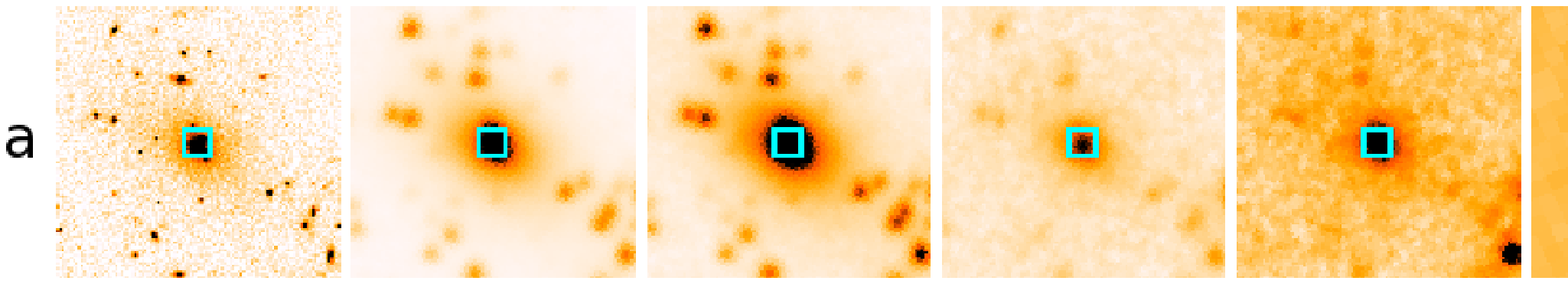}
\includegraphics[angle=0,scale=0.37]{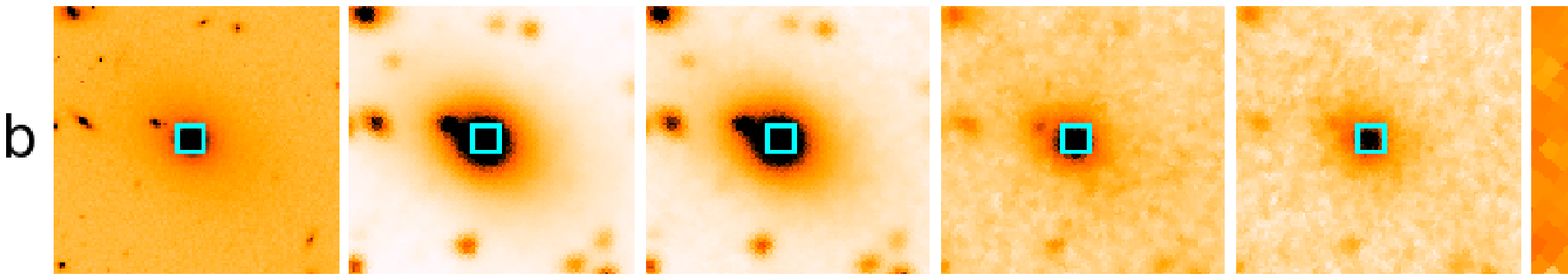}
\includegraphics[angle=0,scale=0.37]{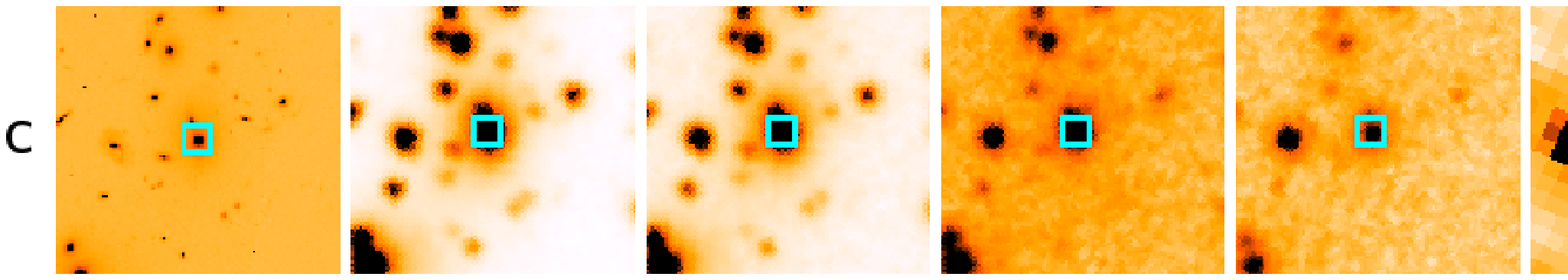}
\includegraphics[angle=0,scale=0.37]{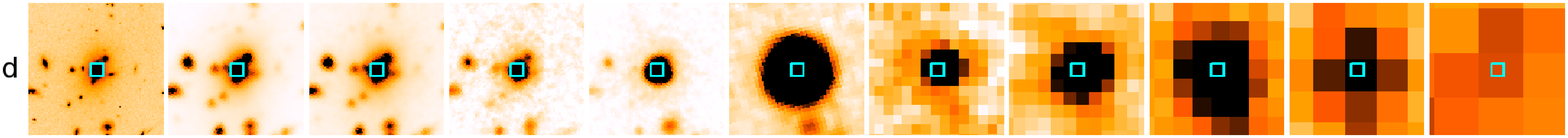}
\includegraphics[angle=0,scale=0.37]{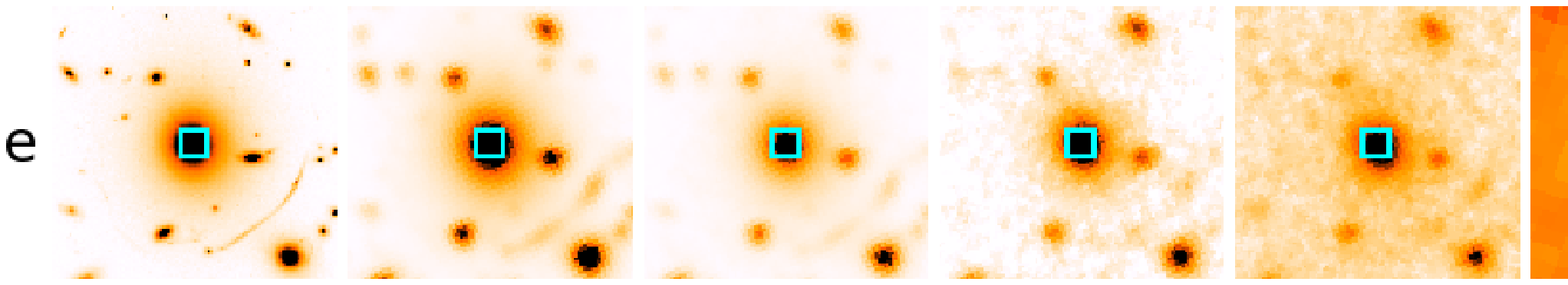}
\includegraphics[angle=0,scale=0.37]{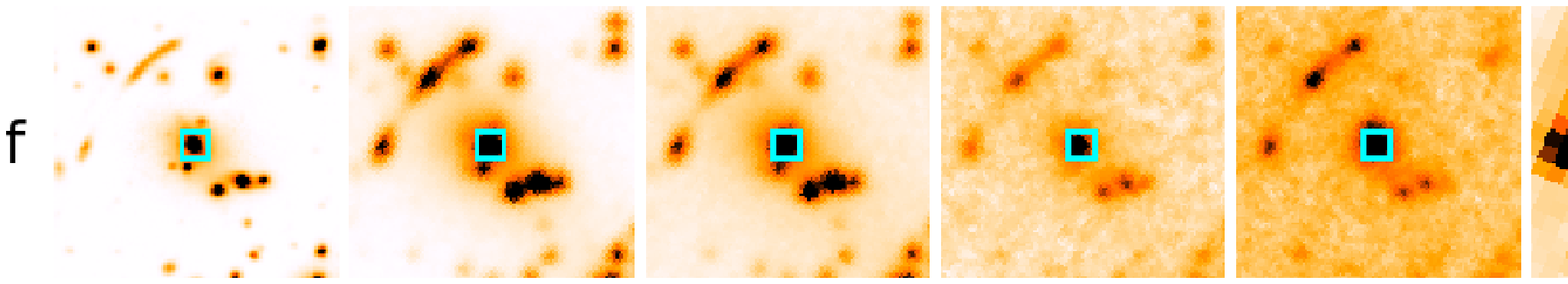}
\caption{11-band thumbnails (HST, IRAC, MIPS, PACS, SPIRE) for a representative selection of BCGs, identified by the cyan square (each frame is 40"$\times$40"): {\it (a)} AS1063, {\it (b)} A1914, {\it (c)} A2744, {\it (d)} A1835, {\it (e)} A383, {\it (f)} A2667. See Section \ref{sec:parts} for details.}
\label{fig:thumbs}
\end{figure*}

\placefigure{fig:thumbs}

\subsection{Counterparts}
\label{sec:parts}

The PACS beam size ($\sim$8 arcsec) has a large probability of encompassing multiple optical sources. The most effective method of evaluating whether long-wavelength flux is associated with an optical source is to `step through' the wavelength range using all available near- and mid-infrared data. It is important to note that {\it a priori} knowledge or expectation of whether a particular BCG should emit in the far-infrared {\bf was not} used during the allocation of flux to sources.

In all cases, the BCG is identifiable in the 2MASS and {\it Spitzer} IRAC near-infrared bands, when such data is available (only 7 BCGs lack IRAC coverage), providing a useful link between optical and far-infrared wavelengths. In many cases, there are no {\it Spitzer} 24 \micron{} or Herschel sources near the optical BCG position, and the BCG clearly decreases in luminosity through the IRAC bands (e.g. Figure \ref{fig:thumbs}{\it a--b}). In a few instances, a single mid-infrared source is located in the immediate vicinity of the optical BCG position, but on closer examination, this flux obviously originates from a secondary IRAC source (e.g. Figure \ref{fig:thumbs}{\it c}).

A few clusters contain single, point-source FIR counterparts to the BCG (e.g. Figure \ref{fig:thumbs}{\it d}). However, many of the Herschel-detected BCGs require the subtraction of secondary sources in at least the longer SPIRE bands where the spatial resolution deteriorates (e.g. Figure \ref{fig:thumbs}{\it e--f}). The techniques employed for extracting flux, particularly in such blended and confused cases, are described in the next section.

\subsection{Photometry}

Photometry for {\it Spitzer} and PACS, where blending and source confusion is minimal, was obtained using simple aperture photometry via the {\sc SExtractor} package \citep{ber96-393}. Aperture corrections were applied individually in each band, based on instrument handbook values or the latest model beam profiles.

In the SPIRE bands, background source confusion is above the instrumental noise. Photometry for all sources within $\sim$3 arcmin of the BCG was extracted using the {\sc Iraf} routine {\sc Allstar}. This algorithm fits the known PSF to all positions in an input source catalog, thereby measuring photometry for all potential sources simultaneously. Any significant contaminating flux within a beam of the BCG position should be accounted for. Where available, MIPS 24 \micron{} is the favored input catalog, as the majority of Herschel point sources have visible counterparts in the MIPS image. Otherwise, the longest available wavelength IRAC band is used instead (although the correspondence between IRAC and FIR is not as consistent). While residual flux from confused background sources, too faint for inclusion in the MIPS catalog, may boost the SPIRE flux, use of such an {\it a priori} catalog should minimize the effect. Statistical de-boosting is beyond the scope of this study.

Table \ref{tab:phot} presents the Herschel fluxes (100-500 \micron) for the HLS and LoCuSS sample BCGs. Non-detections are indicated by the 3$\,\sigma$ detection limit of each map (note the significant difference in depth between HLS and LoCuSS data). Overall, 37 of the 65 BCGs in HLS+LoCuSS are detected at 24 \micron{}. A further 7 BCGs lack observations at this wavelength. However, in all such cases the slope of the IRAC photometry together with non-detections in PACS, suggest no significant infrared component (SFR $<$ 2 M$_\sun$yr$^{-1}$). Of the 37 MIPS sources, 12 are detected in both PACS bands (there are no sources in a single Herschel band only). Ten of these BCGs are also detected by SPIRE at 250 \micron{}, 7 at 350 \micron, while two have significant flux at 500 \micron.

\section{SED analysis and results}
\label{sec:results}

In this section we analyze the full sample of 68 BCGs (65 HLS+LoCuSS; 3 from E10), ensuring a homogeneous characterization of the far-infrared component.

\subsection{Infrared SED fitting}
\label{sec:sed}

\begin{figure}
\includegraphics[angle=270,scale=0.58]{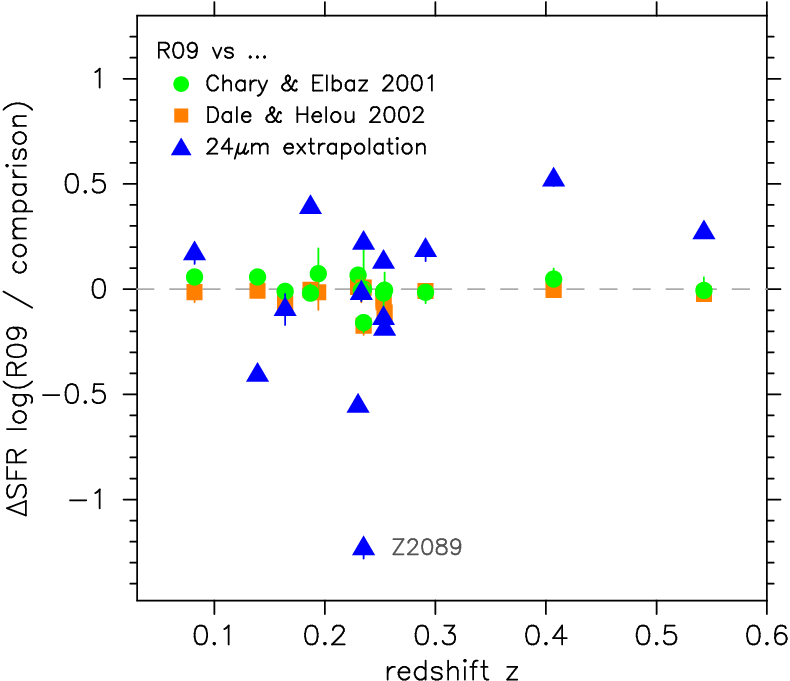}
\caption{For FIR-bright BCGs, obscured SFR derived using the \citet[][R09]{rie09-556} templates, compared to SFRs derived from \citet[][green circles]{cha01-562} and \citet[][orange squares]{dal02-159} templates, and from extrapolating the 24 \micron{} flux via equation 14 of \citeauthor{rie09-556} (blue triangles). All $\Delta$SFRs are plotted as a function of BCG redshift, with the majority of errors well within the symbol size. The labeled major outlier (SFR$_{\rm 24}$ $\gg$ SFR$_{\rm R09}$) is Z2089.}
\label{fig:check}
\end{figure}

\placefigure{fig:check}

Far-infrared dust properties are derived from the best-fitting template across all photometric data $\lambda_{\rm obs}$ $\ge$ 24 \micron{}. We employ weighted fits to the recent templates from \citet[][R09]{rie09-556}. These templates are derived from eleven star-formation-dominated local (ultra--)luminous IR galaxies (LIRGs) with high quality data across the infrared wavelengths, which were combined and interpolated to produce a library of spectra for galaxies at specific luminosity. Recently, inconsistencies have been discovered between the shape of the FIR component in these local sources and more distant galaxy populations. For instance, luminous sources at high redshift appear to be fit better by local templates of lower luminosity \citep[e.g.][]{rex10-13}, while a subset of cluster member galaxies have warmer dust than expected for their luminosity, and are better fit by higher luminosity templates (\citealt{raw10-14}; Rawle et al., in prep). Here, we are primarily interested in the total infrared luminosity, and so disregard the nominal luminosity class of the templates, simply integrating the template which best fits the observed shape of the dust component (with appropriate normalization).

Total infrared luminosity ($L_{\rm TIR}$) is calculated by integrating the best fit template over the (rest frame) wavelength range $\lambda_{\rm 0}$ $=$ 8--1000 \micron. Uncertainty in luminosity is estimated by repeating the full fitting procedure for many Monte Carlo realizations of the observed photometric data. Star formation rate (SFR) is derived directly from $L_{\rm TIR}$ following the simple relation of \citet{ken98-189}\footnote{SFR$_{\rm IR}$ [$M_{\sun}$ yr$^{-1}$] $=$ 4.5$\times$10$^{-44}$ $L_{\rm IR}$ [erg s$^{-1}$]}.  Figures \ref{fig:seds1}--\ref{fig:seds3} in Appendix \ref{app:seds} present the observed FIR SED for Herschel-detected BCGs, and indicates the best fit templates and subsequent values for $L_{\rm TIR}$ and SFR.

As a sanity check, $L_{\rm TIR}$ and SFR are also derived using the templates from \citet{cha01-562} and \citet{dal02-159}. The best fitting template from both are selected as described above for R09. SFRs derived from these template sets agree with the R09 values within 6\% and 2\% respectively (Figure \ref{fig:check}). All three template libraries are unable to match the observed FIR and 24 \micron{} simultaneously for a significant number of sources. This is most likely due to PAH and/or silicate absorption features within the MIPS 24 \micron{} band (rest frame $\lambda_{\rm 0}$ $\approx$ 16--20 \micron{} at these redshifts). SFR can be estimated directly from the 24 \micron{} flux, following equation 14 from \citet{rie09-556}. SFR$_{\rm 24}$ disagrees with the Herschel-derived SFR (from integrated $L_{\rm TIR}$) by an average of 20\% (Figure \ref{fig:check}). The major outlier (Z2089; SFR $\sim$ 0.05$\times$SFR$_{\rm 24}$) is described further in subsequent sections.

Characteristic dust temperature is calculated via a modified blackbody of the form
\begin{equation}
S_{\nu} = N{\nu}^{\beta}B_{\nu}(T_{\rm dust})
\end{equation}

\noindent{}where $S_{\nu}$ is flux density, $\beta$ is dust emissivity index (fixed at 1.5) and $B_{\nu}(T_{\rm dust})$ is the Planck blackbody radiation function for a source at characteristic temperature $T_{\rm dust}$. $\beta$$T_{\rm dust}$ remains approximately constant \citep{bla03-733}, so fixing $\beta$=2.0 instead, would systematically lower the derived temperature (by $\sim$3 K at 30 K). Photometry at wavelengths shorter than 100 \micron{} are not used in the blackbody fit.

Fits for two modified blackbodies, as employed in E10, are strongly dependent on the inclusion of 24 \micron{}. With the mid-infrared, best fit temperatures tend towards $\sim$20 K and $\sim$50 K, with much the same $L_{\rm TIR}$ as the best fitting template. Excluding 24 \micron{}, the fits are less constrained and the two components become more similar ($\sim$30--40 K), or one component becomes insignificant and the solution tends towards the single modified blackbody. At most, only six of our BCG sample (including all three from E10) have sufficient photometry to fit a well constrained two-component model. We avoid two component blackbodies in favor of the single characteristic temperature and $L_{\rm TIR}$ derived from the best fitting template.

Template and blackbody fitting is only employed for BCGs with a detection in at least two Herschel bands (although in practice, no sources have a detection in only one band). For the remaining BCGs, an upper limit on $L_{\rm TIR}$, and hence SFR, can be calculated from the Herschel detection limits. The detection limits are significantly different for each survey (but relatively stable map-to-map within a sub-sample, e.g. 1$\,\sigma$ deviation in the HLS limits is 0.1 and 0.3 mJy for the two PACS bands), and the limit for $L_{\rm TIR}$ is also a function of redshift. An additional limit on luminosity is estimated from the 24 \micron{} data (where available), and is used in subsequent figures if it presents a tighter constraint than the Herschel-derived limit.

\subsection{Total infrared luminosities}
\label{sec:bcgs}

\begin{figure*}
\centering
\includegraphics[angle=270,scale=0.9]{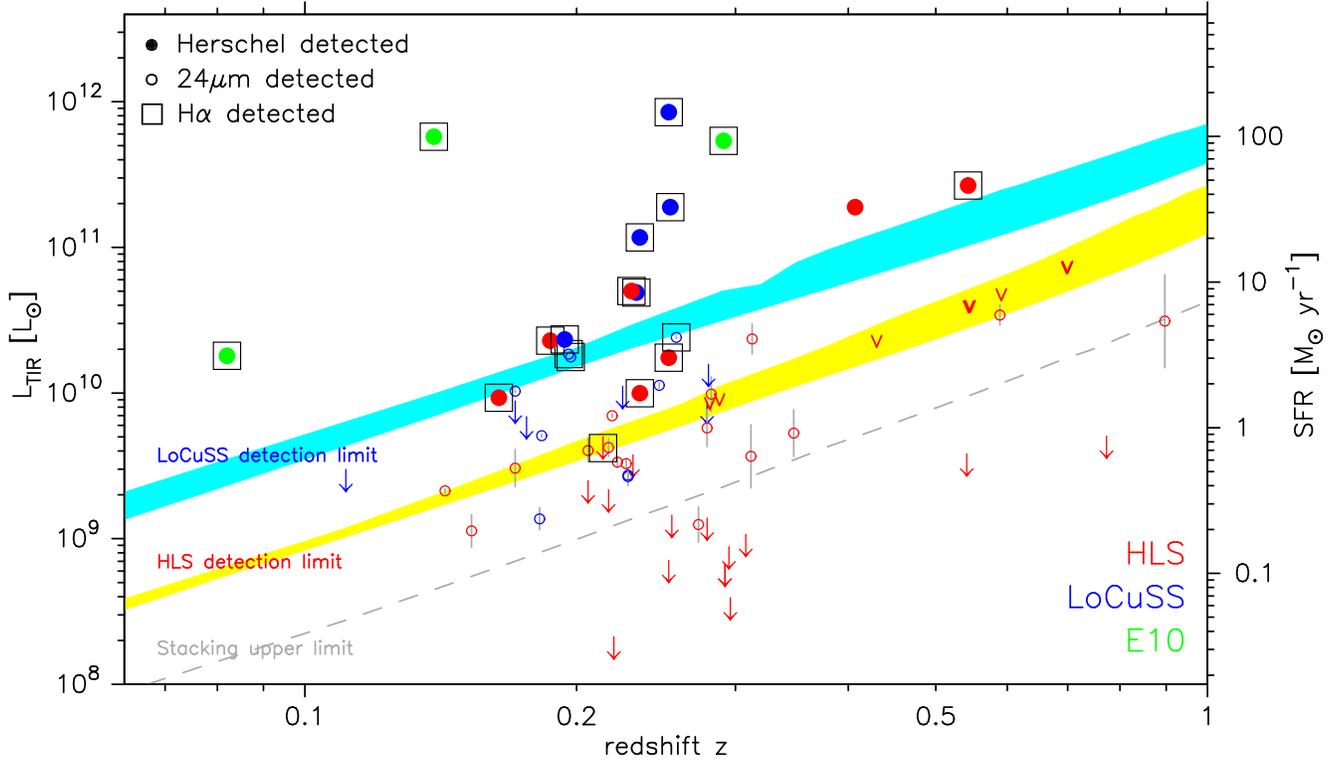}
\caption{Total infrared luminosity, $L_{\rm TIR}$, versus redshift for each BCG: red $=$ HLS; blue $=$ LoCuSS; green $=$ E10. (Red if both HLS and LoCuSS, as HLS data used). Large filled circles denote sources detected by Herschel. Smaller open circles show $L_{\rm TIR}$ estimated from 24 \micron{} (for Herschel non-detections). `v'-shape and arrow symbols indicate sources undetected by Herschel: the former having $L_{\rm TIR}$ upper limits calculated from Herschel detection limits, while the latter have tighter constraints derived from 24 \micron{}. Note that all points have associated errors, but the majority lie within the symbol size. BCGs with detected H$\alpha$ emission are highlighted with an open square, and there is generally a good correspondence between FIR and H$\alpha$ detections (see Section \ref{sec:halpha}). Shaded areas represent the luminosity limit (as a function of redshift) based on the nominal depth of the Herschel data: yellow $=$ HLS; cyan $=$ LoCuSS. The width of these areas results from scatter in individual map detection limits, and the variation in templates of different luminosity from \citet{rie09-556}. The grey dashed line shows the mean upper limit on $L_{\rm TIR}$ of BCGs without H$\alpha$ emission, as derived from the stacking analysis (see Section \ref{sec:stack} for details).}
\label{fig:lumz}
\end{figure*}

\placefigure{fig:lumz}

The upper portion of Table \ref{tab:seds} presents $L_{\rm TIR}$ and SFR (as derived from the R09 fits) for all Herschel-detected BCGs in the sample. The 24 \micron-derived values are shown for completeness. The majority of BCGs in this sample are not luminous in the far-infrared (Figure \ref{fig:lumz}). Only 15 of the 68 BCGs (22$^{+6.2}_{-5.3}$\%) are confirmed as $L_{\rm TIR}$ $\ga$ 10$^{10}$ L$_\sun$ (SFR $\ga$ 2 M$_\sun$yr$^{-1}$), all in the redshift range 0.05 $<$ $z$ $<$ 0.6. The absolute fraction agrees approximately with previous studies using {\it Spitzer} mid-infrared data alone: 27$^{+20}_{-14}$\% from \citet{ega06-922} and 43$^{+8}_{-7}$\% from \citet{ode08-1035}. The difference likely results from the Herschel detection limit of our data compared to the 24 \micron{} sensitivity, and from the H$\alpha$ selection used by \citet{qui08-39} and \citeauthor{ode08-1035}, which implicitly favors cool-core clusters. Seven BCGs (10$^{+5.4}_{-3.4}$\%) are LIRGs (i.e. $L_{\rm TIR}$ $>$ 10$^{11}$ L$_\sun$; SFR $\ga$ 20 M$_\sun$yr$^{-1}$), which is again similar to, but marginally smaller than, the 15$\pm$5\% from \citeauthor{ode08-1035}. The brightest BCG in our sample is A1835, with an $L_{\rm TIR}$ $=$ 8.5$\times$10$^{11}$ L$_\sun$.

The combined cluster sample is in no sense complete, with a bias towards the most massive and well studied clusters (primary targets for lensing) and containing three BCGs specifically selected due to their anticipated high FIR luminosity. The 31 LoCuSS clusters can be considered statistically identical to a volume limited sample selected on X-ray luminosity only (see \citealt{smi10-18} for more details). The clusters within LoCuSS (including the 11 overlapping with HLS), contain 32 BCGs, eight of which are detected in the far-infrared. This fractional value (25$^{+10.0}_{-8.2}$\%) is very similar to our full BCG sample. However, an additional 3 LoCuSS BCGs have a 24 \micron{}-estimated $L_{\rm TIR}$ $>$ 10$^{10}$ L$_\sun$, but fall below the Herschel detection limit. Including these, the fraction of star-forming BCGs in the LoCuSS sample (SFR $>$ 2 M$_\sun$yr$^{-1}$) is 34$^{+10.4}_{-9.7}$\%.

The nominal detection limits for each sample are also illustrated in Figure \ref{fig:lumz}, and they effectively vary with redshift as a power law, meaning the observations probe $\sim$2.5 dex in $L_{\rm TIR}$ deeper for the closest clusters ($z$$\sim$0.1) than the most distant ($z$$\sim$1.0). The Herschel observations for HLS probe $L_{\rm TIR}$ $>$ 10$^{10}$ L$_\sun$ for all clusters $z$ $\la$ 0.35. The variation of the detection limit, combined with the small number of detections and over-abundance of sources at $z$ $\sim$ 0.2--0.3 (due to the LoCuSS selection criterion) makes it difficult to comment on possible evolution over time.

\subsection{AGN contamination}
\label{sec:agn}

Near- and mid-infrared studies have reported that IR-bright BCGs show decreasing flux through the IRAC bands, as expected for a stellar dominated SED (e.g. \citealt{ega06-922}). In contrast, an AGN-dominated source would exhibit a power-law increase through the near-infrared. The majority of BCGs in this present sample also appear to be dominated by the stellar component, and $L_{\rm TIR}$ estimated from 24 \micron{} does not contradict the Herschel detection limits. MS2137 ($z$$=$0.31) is an exception to this trend, with a luminosity estimate from 24 \micron{} $\sim$0.5 dex higher than the nominal detection limit of the Herschel data. The IRAC bands suggest this is not due to a dominant AGN. We return to this sources in Section \ref{sec:halpha}.

Z2089 is the one BCG to show a clear power-law increase from $\sim$5--70 \micron{} (in IRAC/MIPS and WISE photometry). The IRS spectrum (Egami et al., in prep.) confirms a power-law continuum increase for 5--35 \micron{}, while optical spectroscopy (MMT Hectospec; Pereira et al., in prep.) shows broad emission and line ratios suggestive of a strong AGN. The SED for Z2089 shows a large discrepancy between MIPS 70 \micron{} and the Herschel observations (almost a magnitude lower in Herschel; see Figure \ref{fig:seds2}). IRAC, WISE and 24 \micron{} are tracking the AGN component continuum seen in the IRS spectrum (SFR$_{24}$ $=$ 432 M$_\sun$ yr$^{-1}$ without accounting for an AGN). However, it is likely that 70 \micron{} is substantially boosted by bright [OIII], [NIII] and/or [OI] emission lines (at $\lambda_{\rm rest}$ $=$ 51.8, 57.3, 63.3 respectively) which lie within the 70 \micron{} band at $z$$\sim$0.24. An alternate explanation is that the notoriously shallow {\it Spizer} 70 \micron{} imaging (in this case from \citealt{qui08-39}) has given an erroneous detection. Even if that were the case, the FIR peak would have to be particularly broad, or blue, to encompass the WISE/24 \micron{} and PACS data. This highly discrepant 70 \micron{} point makes it difficult to include {\it Spitzer} data in the SED fitting procedures, and the AGN component is much less dominant beyond 100 \micron{}. Therefore, our Herschel-derived $L_{\rm TIR}$ ($\sim$20 M$_\sun$ yr$^{-1}$) is a very poor fit $\lambda_{\rm obs}$ $\le$ 70 \micron{}, but is likely to be representative of the true obscured SFR (and is at least a much tighter upper limit).

Optical emission line ratios indicate that A1835 and to a lesser extent A1068, may have sub-dominant contributions from AGN \citep{qui08-39,don11-40}. In contrast to Z2089, these AGN components are not identifiable from IRAC colors, the star-formation dominated templates fit well at longer wavelengths, and the dust temperatures are reasonable given the FIR luminosity. SFR$_{\rm FIR}$ is unlikely to be boosted significantly. 

\subsection{General dust properties}

\begin{figure*}
\centering
\includegraphics[angle=270,scale=0.65]{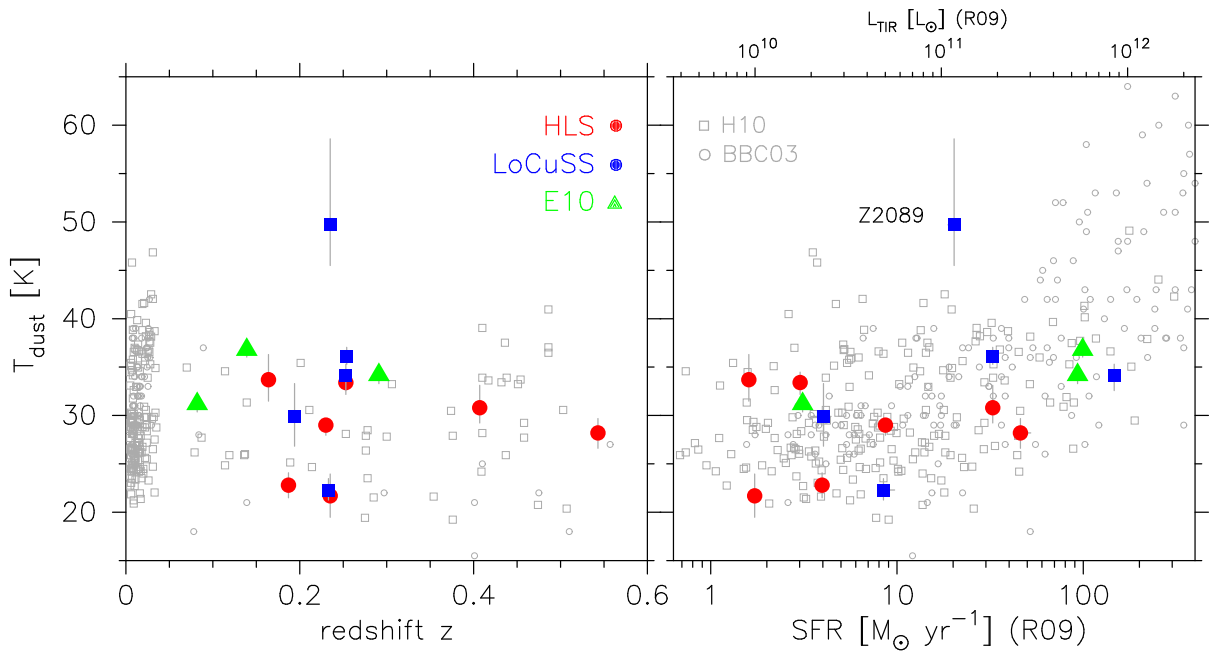}
\caption{Characteristic dust temperature, as derived from a single modified blackbody fit to the Herschel observations, as a function of redshift ({\it left panel}) and SFR ({\it right panel}) for the FIR-bright BCG sample (large colored points). In the right-hand panel, the equivalent total infrared luminosity is also shown for reference. The normal star-forming galaxy samples from \citet{bla03-733} and \citet{hwa10-75} are plotted as open grey symbols for comparison: in the left panel, restricted to $L_{\rm TIR}$ $<$ 3$\times$10$^{11}$ L$_\sun$ as for a typical BCG; in the right panel restricted to $z$ $<$ 0.6. The majority of these star-forming sources are in the local universe ($z$ $<$ 0.05). The exceptionally warm source is the AGN-dominated BCG Z2089.}
\label{fig:tbb_plot}
\end{figure*}

\placefigure{fig:tbb_plot}

We briefly explore the general dust characteristics of the FIR-bright BCG population. The modified blackbody fit to far-infrared data reveals a typical range of dust temperatures 20--40 K (Figure \ref{fig:tbb_plot}). As expected for a source with a strong AGN component, the dust temperature of Z2089 is elevated relative to the normal star-forming population. Excluding Z2089, and accounting for the small sample size, there is no apparent evolution of BCG dust temperature with redshift. There is a wide range in star formation rates, with the three most active attaining SFR $\ga$ 100 M$_\sun$ yr$^{-1}$, although as previously stated, two of these may include a modest AGN contribution. As expected from simple models of dust heating, and confirmed by almost a decade of observations \citep[e.g.][]{bla03-733,hwa10-75}, the most actively star-forming galaxies typically have warmer dust (grey symbols in Figure \ref{fig:tbb_plot}). In our BCG sample, the LIRG-type BCGs ($L_{\rm TIR}$ $>$ 10$^{11}$ L$_\sun$) are on average $\sim$5 K warmer than the sub-LIRGs. Generally, BCGs (other than the IR-bright AGN host, Z2089) are indistinguishable from normal star-forming galaxies in this parameter space.

\section{Discussion}
\label{sec:disc}

\subsection{Comparison to cool-core diagnostics}
\label{sec:halpha}

Separation of cool-core and non-cool-core clusters can be difficult and often contentious. The X-ray cooling time is a clean primary indicator of cool gas in the cluster center \citep{hud10-37}. We follow the relation from \citet{don05-13} for pure free--free cooling

\begin{equation}
t_{\rm c0}(K_0) \sim 10^8 {\rm yrs} \left( \frac{K_0}{10\, {\rm keV\, cm}^2} \right)^{3/2} \left( \frac{kT_X}{5\, {\rm keV}} \right)^{-1}
\end{equation}

\begin{figure*}
\centering
\includegraphics[angle=270,scale=0.93]{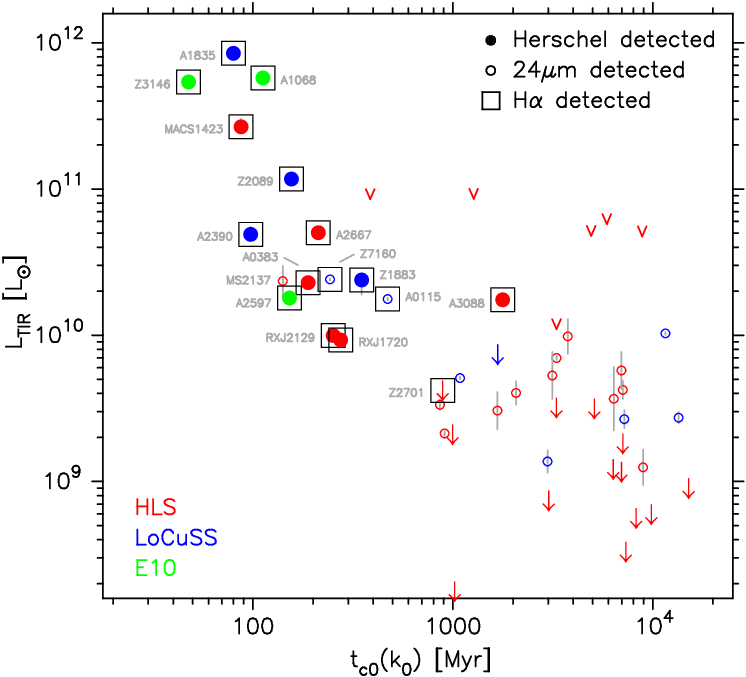}
\caption{$L_{\rm TIR}$ (from R09 templates) versus the central X-ray cooling time $t_{\rm c0}(K_0)$ derived from the ACCEPT Chandra archive \citep{cav09-12} entropy profiles. Solid circles are Herschel detections; open circles are 24 \micron{} detected only;  `v' and arrows for upper-limits from Herschel and 24 \micron{} respectively. Colors as in Figure \ref{fig:lumz}. BCGs with detected H$\alpha$ emission are highlighted with an open square and labeled. For $t_{\rm c0}(K_0)$ $<$ 1 Gyr, there is a very good correlation between IR luminosity and cooling time.}
\label{fig:cooling}
\end{figure*}

\noindent{}where $K_0$ is the central entropy derived from a simple power law profile, and $kT_X$ is the mean X-ray gas temperature. The ACCEPT Chandra archival survey \citep{cav09-12} presents both $K_0$ and $kT_X$ for 50 of the 66 clusters in our sample (including 12 of 14 containing FIR-bright BCGs). Figure \ref{fig:cooling} compares $L_{\rm TIR}$ to this cooling time, and shows a very good anti-correlation for the Herschel detected BCGs ($t_{\rm c0}$ $\la$ 1 Gyr). This confirms the relation (based on 24 \micron{}-extrapolated $L_{\rm TIR}$) presented in \citet[][figure 9]{ode08-1035}, and strongly suggests that the obscured star formation in BCGs is connected to the cluster-scale cooling process.

Ideally we would like to compare the rate of mass deposited ($\dot{M}$) by the cooling central ICM to the condensation rate as traced by SFR. For a small sample of {\it Spitzer}-detected BCGs, \citet{ode08-1035} show (in their figure 10) that $\dot{M}$ is $\sim$0.5--2 orders of magnitude higher than the SFR. This suggests that the cooling ICM is a plausible origin for the star formation fuel. Computing homogeneous central mass deposition rates for our entire cluster sample (based on e.g. ACCEPT data) remains a priority, but is beyond the scope of this current paper.

Section \ref{sec:agn} identifies three of the most IR-luminous BCGs as possible hosts of mid-infrared bright AGN. One may worry that they are so FIR-bright purely because of the contribution of the AGN heating. However, these BCGs are also located in clusters with very short cooling times ($<$ 200 Myr), and if the correlation in Figure \ref{fig:cooling} is real, then they would be expected to be amongst the most highly star-forming. The apparent correlation (albeit very small number statistics) between the BCGs with the highest SFR and those with the strongest radiative AGN may be further evidence \citep[e.g.][]{ega06-922} that substantial cooling of intracluster gas in the cluster core modifies the state of the AGN, part of the cyclic interplay between cooling and heating within AGN feedback \citep[e.g.][]{sil98-1,bow06-645}.

\placefigure{fig:cooling}

We now turn to the cold molecular gas itself, directly detectable using millimeter spectroscopy \citep[e.g.][]{edg01-762,sal03-657}. However, even with the advent of Herschel spectroscopy in the far-infrared \citep{edg10-46}, sensitivity severely restricts the sample of potential BCG targets. In this analysis, we opt to use H$\alpha$ line emission, a known and reliable indicator of $\sim$10$^4$ K molecular gas \citep[e.g.][]{cra99-857,cav08-107}. The H$\alpha$ observations are from a mix of imaging and spectroscopic observations: \citet{don92-49}, \citet{cra99-857}, \citet{qui08-39}, the Sloan Digital Sky Survey \citep{ade08-297} and the REFLEX BCG sample (PI: Edge).

Eighteen of the 68 BCGs (26$^{+6.4}_{-5.7}$\%) have detected H$\alpha$ line emission (highlighted by squares in Figures \ref{fig:lumz} and \ref{fig:cooling}), agreeing well with the absolute fraction reported by previous large H$\alpha$ surveys \citep[e.g. 27$\pm$4\% for 203 BCGs from][]{cra99-857}. The generally good correspondence between Herschel-detection and those with measured H$\alpha$ emission (78\% of H$\alpha$-detections are FIR sources, while 93\% of FIR sources have H$\alpha$ detections; Figures \ref{fig:lumz}) suggests that the two measurements are probing the same phenomenon, and are reaching a comparable depth. Four H$\alpha$ detections originate from BCGs undetected by Herschel (the properties of these are listed below the dividing line in Table \ref{tab:seds}). Three of these sources are covered by the shallower LoCuSS far-infrared observations, and $L_{\rm TIR}$(24 \micron) indicates that they are only marginally below the Herschel detection limit. The remaining BCG in this category (Z2701) has a very low $L({\rm H}\alpha)$ ($>$0.6 dex below any other detection). The H$\alpha$ sources undetected by Herschel do not, therefore, contradict a FIR--H$\alpha$ connection. Only one BCG (A851) is detected by Herschel but not in H$\alpha$; we return to this source in the final paragraph of the section.

There is a very good correspondence between H$\alpha$ detection and short cooling times (Figure \ref{fig:cooling}). Almost all BCGs in clusters with $t_{\rm c0}(K_0)$ $<$ 1 Gyr are detected in H$\alpha$, confirming that the emission is indeed a good indicator of a cool-core cluster \citep[as used in e.g.][]{cav08-107,mcd11-33}. The one notable exception is MS2137, which is not detected by either Herschel or H$\alpha$, yet has $t_{\rm c0}(K_0)$ $\sim$ 140 Myr. This cooling time fits remarkably well with the 24 \micron-estimated $L_{\rm TIR}$, but the Herschel non-detection shows that such a luminosity is not possible (the limit from Herschel is $\sim$0.5 dex lower). The BCG is coincident with the X-ray peak, so if it were conforming to the general trend, the cold intracluster medium should be feeding star formation in the galaxy. Further explanation of this BCG requires additional study beyond this current paper.

\begin{figure*}
\centering
\includegraphics[angle=270,scale=1.1]{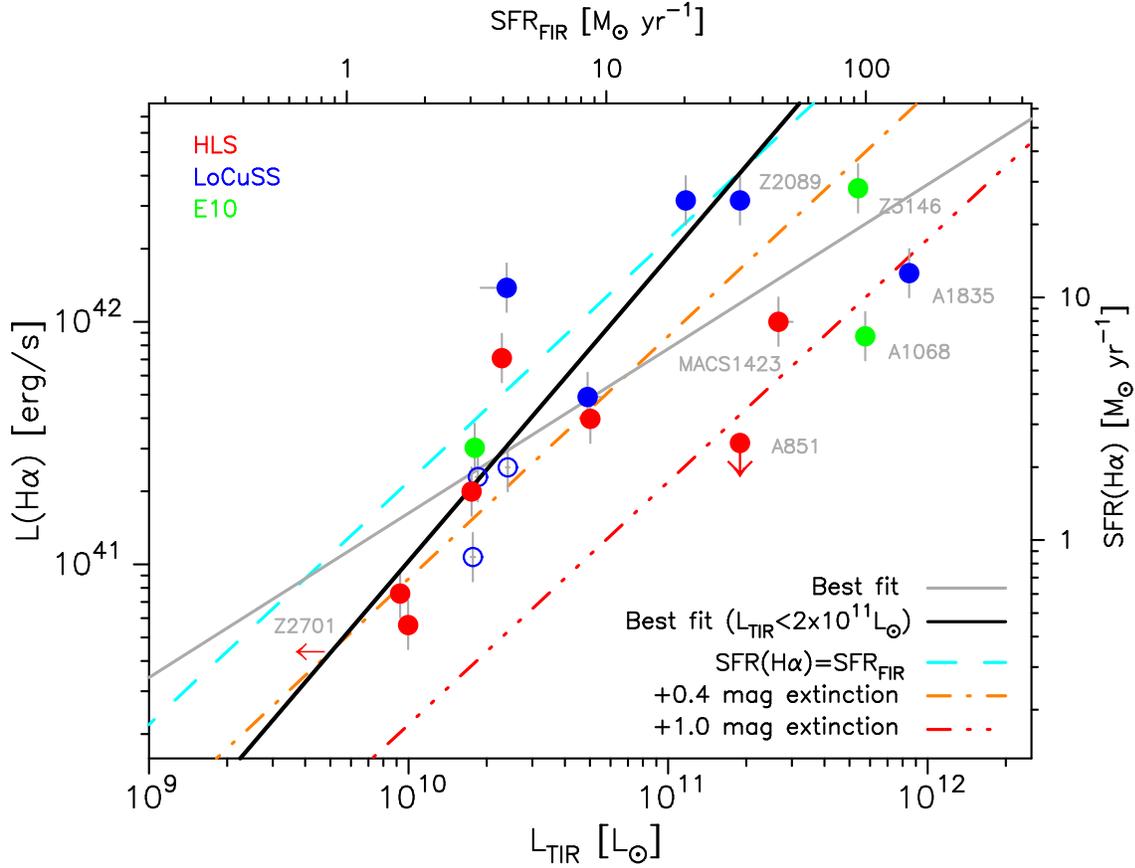}
\caption{$L({\rm H}\alpha)$ versus $L_{\rm TIR}$ (from R09 templates) for BCGs included in Table \ref{tab:seds}. Solid circles are Herschel detections; open circles are 24 \micron{} detected only. Colors as in Figure \ref{fig:lumz}. Selected BCGs are labeled (see text for further details). Grey solid line shows the best fit to all BCGs with detections in both Herschel and H$\alpha$ (rms scatter $=$ 0.34 dex), while the black solid line is the best fit to those $L_{\rm TIR}$ $<$ 2$\times$10$^{11}$ $L_{\odot}$ (scatter $=$ 0.27 dex). The cyan dashed line represents the relation log[$L({\rm H}\alpha)$] $=$ log($L_{\rm TIR}$) $+$ 31.34 (i.e. SFR(${\rm H}\alpha$)$=$SFR$_{\rm FIR}$ via \citealt{ken98-189}). The orange dash--dot and red dash-dotted lines show the same relation, but accounting for 0.4 and 1 mag respectively of uncorrected extinction in H$\alpha$ flux. Corresponding SFRs, via \citet{ken98-189} are also displayed.}
\label{fig:halpha}
\end{figure*}

We now qualitatively compare $L({\rm H}\alpha)$ and the total infrared luminosity (Figure \ref{fig:halpha}). We have endeavored to remove inconsistencies in $L({\rm H}\alpha)$, such as different aperture sizes and $H_0$ assumptions (for the older literature measurements). The values are corrected for extinction from our Galaxy, but not for internal reddening, and treatment of the [NII] line is different in the various surveys (see individual papers for more details). Despite these uncertainties and caveats, there appears to be a reasonable correlation ($\sim$0.34 dec scatter; grey solid line) over two orders of magnitude (in $L_{\rm TIR}$). The four H$\alpha$-emitting BCGs undetected by Herschel but with $L_{\rm TIR}$(24 \micron), lie comfortably on this trend. On closer examination, there appears to be a dichotomy between the highest $L_{\rm TIR}$ sources and the others. Fitting to only BCGs $L_{\rm TIR}$ $<$ 2$\times$10$^{11}$ $L_{\odot}$, the slope steepens significantly (black solid line), and the scatter decreases to 0.27 dex. 

\placefigure{fig:halpha}

Normal star-forming galaxies typically exhibit extinction $A({\rm H}\alpha)$ $\sim$ 1 mag \citep[e.g.][]{ken83-54,hop01-288}. For more IR-luminous galaxies ($L_{\rm TIR}$ $>$ 10$^{11}$ $L_{\odot}$), dust may become increasingly opaque to optical wavelengths and the total SFR tends towards SFR$_{\rm FIR}$ \citep{cal10-1256}. From the self-consistent relations of \citet{ken98-189}\footnote{SFR$_{\rm FIR}$ [M$_\sun$ yr$^{-1}$] $=$ 1.73$\times$10$^{-10}$$L_{\rm TIR}$ [L$_\sun$];\\ SFR(${\rm H}\alpha$) [M$_\sun$ yr$^{-1}$] $=$ 7.9$\times$10$^{-42}$$L({\rm H}\alpha)$ [erg s$^{-1}$]} zero extinction (i.e. SFR$_{\rm FIR}$ $=$ SFR(H$\alpha$)) would equate to log[$L({\rm H}\alpha)$] $=$ log($L_{\rm TIR}$) $+$ 31.34 (cyan dashed line). The red dash--dotted line in Figure \ref{fig:halpha} indicates 1 mag of (uncorrected) reddening, which could easily account for the highest luminosity BCGs.  However, the best fit line for the lower luminosity BCGs (black solid) lies well above this relation. The orange dash--dot line indicates SFR(H$\alpha$) $=$ SFR$_{\rm FIR}$ -- 0.4 mag uncorrected reddening, suggesting that the majority of the low-luminosity sources can be explained by this relatively small extinction or less. The cold gas appears to be less dusty than in normal star-forming galaxies, indicating a different origin for the fuel.

It is worth noting that the relation indicating a small apparent extinction could also result from an intrinsically larger H$\alpha$ flux, e.g. due to an additional heat source \citep{fab11-172}. For this scenario to be plausible, the effect would need to be more pronounced in the less IR-bright BCGs, which could be true if the absolute contribution to the observed H$\alpha$ emission was similar in all BCGs, and negligible in those with the highest SFRs.

The surprising H$\alpha$ non-detection of A851 suggests a much higher extinction ($A({\rm H}\alpha)$ $\ga$ 1 mag) than the typical BCG. Further examination of the HST optical imaging reveals a large tidal tail, while the BCG also has a projected offset from the X-ray peak of $\sim$280 kpc \citep{bil08-1637}, 10--20$\times$ greater than the majority of this sample. Even if a cool-core were present it  would have little influence. It is likely that galaxy--galaxy interaction (merger or harassment), rather than the cool--core phenomenon, is responsible for the star formation, and hence the extinction resembles the normal star-forming galaxy population.

\subsection{Far-infrared stacking analysis}
\label{sec:stack}

We now attempt to constrain the FIR luminosity of ``non-cool-core" cluster BCGs. To achieve this, the BCGs in the HLS and LoCuSS sub-samples are classified into three exclusive categories: detected in FIR and H$\alpha$, H$\alpha$-detected only (ie not FIR-detected), and undetected in either FIR or H$\alpha$. Note that (1) the three BCGs from E10 would all fall into the first class, and (2) we neglect a FIR-detected only category as it would consist of the debatable BCG in A851 alone. For each class, the Herschel maps are co-added (stacked) after centering on the optical BCG position and removing exceptionally bright nearby sources via point source fitting.

\begin{figure*}
\centering
\includegraphics[angle=0,scale=0.4]{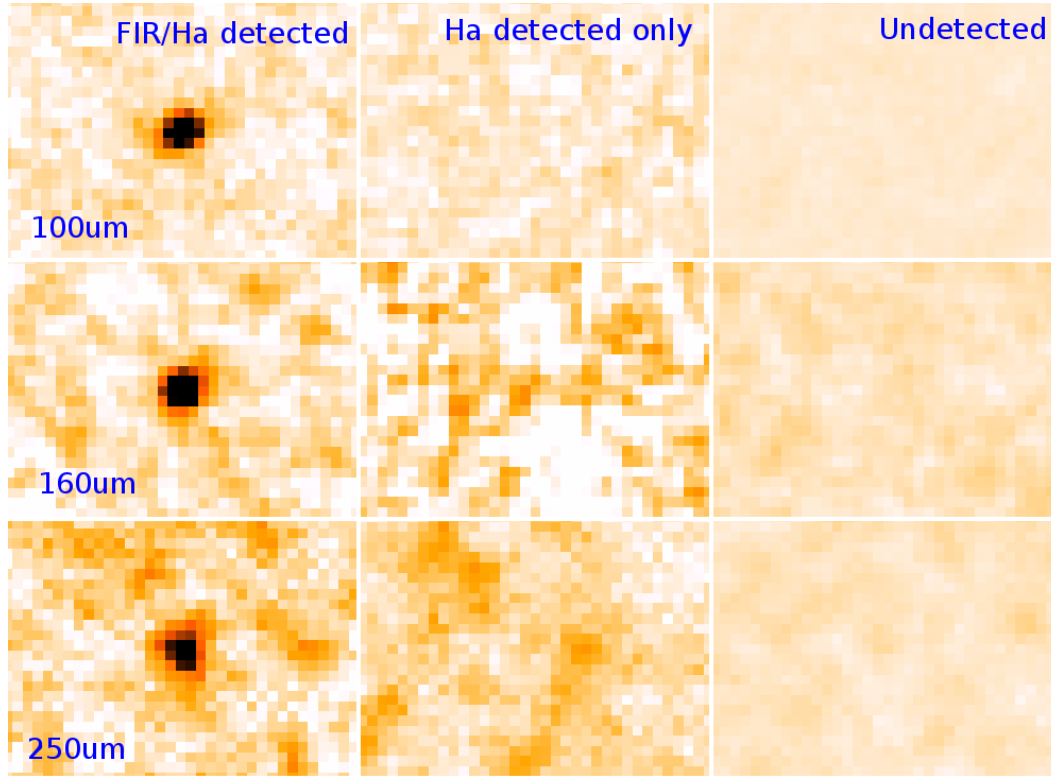}
\caption{Co-added Herschel maps ({\it upper--lower:} 100, 160, 250 \micron) stacked on optical BCG position for the HLS and LoCuSS subsamples. BCGs are categorized by detections in FIR and H$\alpha$ emission. {\it Left panels:} 11 BCGs with both Herschel and H$\alpha$ detections. {\it Center panels:} 4 BCGs detected in H$\alpha$, but not by Herschel. {\it Right panels:} 49 BCGs without FIR or H$\alpha$ detections. No detection is recovered in the latter two classes, with the undetected (non-cool-core cluster) stack revealing mean 3$\,\sigma$ limits of 0.6, 1.1, 1.3, 1.4, 1.8 mJy at 100--500 \micron.}
\label{fig:stacking}
\end{figure*}

Confusion noise is a time-invariant rms uncertainty and hence once reached, integrating longer does not improve the detection limit (in contrast to the more familiar instrument noise). However, confusion noise is still a zero-averaged gaussian rms and is not spatially invariant, so stacking on multiple known source positions from a prior catalog can detect a mean flux below the nominal confusion limit (e.g. \citealt{mar09-1729}). Figure \ref{fig:stacking} presents the PACS and SPIRE 250 \micron{} stacks for each of three categories. For brevity, we analyze but do not show the longer wavelength SPIRE stacks, which are broadly comparable to 250 \micron{}.

As one would expect, the FIR-detected stacks (11 co-added frames) exhibit a strong source in the center (left panels, Figure \ref{fig:stacking}). This is a useful verification of the stacking algorithm and optical--FIR alignment. The aperture corrected flux in the stacks equals (to within $\sim$5\%) the sum of the individual source fluxes, and the PSF is not significantly larger than the nominal beam size for a single map.

The stack of H$\alpha$-detected sources without Herschel flux (center panels, Figure \ref{fig:stacking}) only comprises 4 BCGs (3 from shallow LoCuSS maps), and does not significantly improve upon the HLS detection limit. Therefore, although 24 \micron{} photometry indicates that at least 3 of these sources should be close to the Herschel limit for an individual map, the stack does not reveal a detection.

The final category, for those undetected in both H$\alpha$ and the far-infrared, includes the vast majority of the sample (49 co-added frames; right panels, Figure \ref{fig:stacking}). These can be considered the non-cool-core cluster BCGs. Again, there is no detection, suggesting that a BCG in this class has a mean 3$\,\sigma$ upper limit in each band (100--500 \micron{}) of 0.6, 1.1, 1.3, 1.4, 1.8 mJy. These values are well below the confusion limit of a single source in all bands (e.g. from \citealt{ber10-30,ngu10-5}). The dashed line in Figure \ref{fig:lumz} shows the approximate luminosity limit (as a function of redshift) derived from these Herschel flux limits. At any given redshift, the average FIR luminosity of a non-cool-core BCG is at least an order of magnitude below those in cool-core clusters, suggesting a mean SFR $<$ 0.17 M$_\sun$yr$^{-1}$ at z $\sim$ 0.2 and SFR $<$ 0.42 M$_\sun$yr$^{-1}$ at z $\sim$ 0.3. The stacking analysis supports a bi-modal distribution of far-infrared luminosity, whereby BCGs are either star-forming (due to a cooling ICM), or are as quiescent as normal cluster early-types.

\placefigure{fig:stacking}

\subsection{Origin of the dusty gas}
\label{sec:origin}

We have presented strong circumstantial evidence that the fuel for star formation in FIR-bright BCGs originates from the cooling ICM. A recent study by \citet{voi11-24} asserts that the dusty gas may instead be from normal stellar mass loss, estimating a potential mass contribution up to 8 M$_\sun$yr$^{-1}$. We explore this further by calculating the stellar/dust mass ratio for our BCG sample, including limits on the ratio for the population undetected by Herschel.

Dust masses are estimated at 500 \micron{} via $T_{\rm dust}$ derived from the blackbody fit, using the formula

\begin{equation}
M_{\rm dust} = \frac{4\pi D^2 f_{\rm 500}}{\kappa_{\rm abs} 4\pi B_{\lambda}(T_{\rm dust})}
\end{equation}

\noindent{}where $f_{\rm 500}$ is the rest frame 500 \micron{} flux, $B_{\lambda}$ is the Planck function and $\kappa_{\rm abs}$ $=$ 0.95 cm$^2$ g$^{-1}$ is the absorption coefficient at 500 \micron{} \citep{dra03-241}. Rest frame 500 \micron{} flux is estimated from the best fit R09 template. We derive stellar luminosity from the 2MASS total $K$-band magnitude, adjusted for galactic extinction and K-correction. The redshift-dependent conversion from $K$-band luminosity to stellar mass follows the formulation presented in \citet{arn07-137}\footnote{log(M/L$_K$) $=$ (--0.27 $\pm$ 0.03)$z$ -- (0.05 $\pm$ 0.03)}.

\begin{figure}
\centering
\includegraphics[angle=270,scale=0.6]{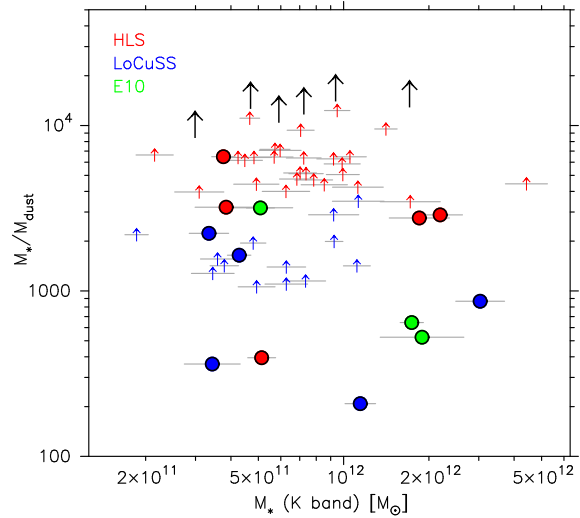}
\caption{Ratio of stellar mass to dust mass, M$_*$/M$_{\rm dust}$, versus the stellar mass. Circular symbols represent Herschel-detected BCGs. Lower limits for Hershel non-detections are given by small colored arrows (colors indicate sample as in previous plots; note HLS galaxies are better constrained as the limits on FIR flux, and hence dust mass, are lower). When FIR-undetected BCGs are stacked within bins of similar stellar mass, the limits in M$_*$/M$_{\rm dust}$ can be raised to the larger black arrows (any stack of individually undetected BCGs gives a non-detection; Figure \ref{fig:stacking}).}
\label{fig:stellar_dust_mass}
\end{figure}

The Herschel-detected BCGs have 200 $<$ M$_*$/M$_{\rm dust}$ $<$ 6000, with a mean M$_*$/M$_{\rm dust}$ $=$ 1570 (Figure \ref{fig:stellar_dust_mass}). The limits for individual non-detected LoCuSS BCGs are inconclusive as the Herschel imaging is fairly shallow, while those from HLS imaging have limits on M$_*$/M$_{\rm dust}$ of 4000--10000. Given that the entire stack of FIR undetected BCGs shows no Herschel flux (Figure \ref{fig:stacking}), we can bin the undetected BCGs by stellar mass (choosing 6 bins with seven BCGs each) and estimate the limit on M$_*$/M$_{\rm dust}$ for the stack. The limit for each stellar mass bin is very similar with a mean M$_*$/M$_{\rm dust}$ $>$ 15000, or an order of magnitude higher than the ratio for Herschel detected BCGs of the same stellar mass. If stellar mass loss were the dominant origin for the dusty gas driving BCG star formation, we would expect all BCGs of a given stellar mass to show similar star formation rates. Otherwise, there would still be a need for a mechanism to trigger star formation in some BCGs while leaving the majority quiescent. We demonstrate a very good relation between SFR and central cooling time in Figure \ref{fig:cooling}, indicating a much stronger probability that the star formation is driven {\bf and} fueled by the cooling ICM.

\section{Summary}
\label{sec:sum}

We present analysis of the largest cluster sample observed with at least 5 Herschel band (100--500 \micron), combining the Herschel Lensing Survey and LoCuSS datasets. In total, the sample comprises 68 BCGs 0.08 $<$ $z$ $<$ 1.0, of which 15 (22$^{+6.2}_{-5.3}$\%) are detected by Herschel, equivalent to $L_{\rm TIR}$ $>$ 10$^{10}$ L$_{\sun}$ (SFR $\ga$ 2 M$_\sun$yr$^{-1}$), including 7 classified as LIRGs. This fractional detection rate is similar to the LoCuSS clusters alone (25$^{+10.0}_{-8.2}$\%), which can be considered statistically identical to a volume-limited X-ray luminosity selected sample. Eighteen of the 68 BCGs (26$^{+6.4}_{-5.7}$\%) have detected H$\alpha$ line emission. Generally, far-infrared and H$\alpha$ detections corresponds well, with 24 \micron{}-based extrapolations to the far-infrared suggesting that most H$\alpha$-detected BCGs unseen by Herschel are only just below the detection limit.

We present four associated strands of circumstantial evidence which strongly suggest that the fuel for star formation in FIR-bright BCGs is not the same as in normal star-forming galaxies, but is directly connected to the central gas cooling in the cluster core.

\begin{enumerate}
\item A very good anti-correlation between Herschel-derived star formation rate and the X-ray cooling time, for BCGs in cool-core clusters ($t_{\rm c0}(K_0)$ $\la$ 1 Gyr);
\item FIR stacks of the Herschel-undetected BCGs (49 sources) show no discernible flux, placing mean 3$\,\sigma$ upper limits of 0.6, 1.1, 1.3, 1.4, 1.8 mJy on the five bands (100--500 \micron), and implying a mean SFR for non-cool-core cluster BCGs of SFR $\ll$ 1 M$_\sun$yr$^{-1}$ (z $<$ 0.4);
\item A surprisingly low H$\alpha$ extinction ($\sim$0.4 mag) for all but the most IR-luminous, Herschel-detected BCGs ($L_{\rm TIR}$ $<$ 2$\times$10$^{11}$ $L_{\odot}$) compared to similar normal star forming-galaxies ($\sim$1 mag; e.g. \citealt{ken83-54}), implying that the gas is significantly less dusty;
\item The stellar-to-dust mass ratio of Herschel-detected BCGs is at least an order of magnitude below the ratio for non-cool-core BCGs of the same stellar mass, suggesting that normal stellar mass loss is unlikely to be the dominant star formation fuel source.
\end{enumerate}

We find no significant evolution in BCG dust properties with redshift, as previously reported in the H$\alpha$ observations of \citet{cra99-857}, and recently shown for all star-forming galaxies in large survey volumes (e.g. \citealt{hwa10-75}). However, we note that our BCG sample is not ideally distributed in redshift space for such an analysis. Despite the difference in H$\alpha$ extinction, the FIR dust properties of BCGs explored briefly in this study (e.g. characteristic temperature) are indistinguishable from the IR-bright population as a whole.

\acknowledgments

This work is partially based on observations made with the Herschel Space Observatory, a European Space Agency Cornerstone Mission with significant participation by NASA. Support for this work was provided by NASA through an award issued by JPL/Caltech. We would also like to thank the HSC and NHSC consortia for support with data reduction. This publication makes use of data products from the Two Micron All Sky Survey, which is a joint project of the University of Massachusetts and the Infrared Processing and Analysis Center/California Institute of Technology, funded by the National Aeronautics and Space Administration and the National Science Foundation. GPS acknowledges support from the Royal Society.

\placetable{tab:phot}

\begin{deluxetable}{lccrcccccc}
\tabletypesize{\small}
\tablecolumns{10}
\tablewidth{0pc}
\tablecaption{Observed Herschel far-infrared (100--500 \micron) photometry for the 65 BCGs in the HLS and LoCuSS samples. Non-detections denoted by the 3$\,\sigma$ limit.}
\tablehead{\colhead{Cluster ID\tablenotemark{a}} & \colhead{$\alpha_{\rm BCG}$\tablenotemark{b}} & \colhead{$\delta_{\rm BCG}$\tablenotemark{b}} & \colhead{$z_{\rm cluster}$} & \colhead{H/L\tablenotemark{c}} & \colhead{$f_{\rm 100}$}  & \colhead{$f_{\rm 160}$} & \colhead{$f_{\rm 250}$} & \colhead{$f_{\rm 350}$} & \colhead{$f_{\rm 500}$} \\ \colhead{} & \colhead{} & \colhead{} & \colhead{} & \colhead{} & \colhead{(mJy)}  & \colhead{(mJy)} & \colhead{(mJy)} & \colhead{(mJy)} & \colhead{(mJy)}}
\startdata
A2744 & 00 14 20.7 & --30 24 00 & 0.308 & H  & $<$2.4 & $<$4.5 & $<$7.8 & $<$9.2 & $<$10.7 \\
MACS0018 & 00 18 33.6 & +16 26 15 & 0.541 & H  & $<$2.3 & $<$4.8 & $<$10.0 & $<$10.4 & $<$11.4 \\
A0068 & 00 37 06.9 & +09 09 24 & 0.255 & HL  & $<$2.5 & $<$5.0 & $<$12.9 & $<$13.0 & $<$13.3 \\
A2813 & 00 43 24.7 & --20 37 41 & 0.292 & H  & $<$2.4 & $<$4.7 & $<$8.3 & $<$10.0 & $<$11.9 \\
A0115 & 00 55 50.6 & +26 24 37 & 0.197 & L  & $<$13.0 & $<$17.0 & $<$14.0 & $<$18.9 & $<$20.4 \\
Z0348 & 01 06 49.4 & +01 03 22 & 0.254 & L  & 81.1 $\pm$  2.8 & 54.7 $\pm$  5.8 & 35.0 $\pm$  5.2 & 12.0 $\pm$  4.3 & $<$20.4 \\
A2895 & 01 18 11.1 & --26 58 12 & 0.227 & H  & $<$2.4 & $<$4.7 & $<$7.7 & $<$9.1 & $<$11.1 \\
A0209 & 01 31 52.5 & --13 36 41 & 0.206 & HL  & $<$2.4 & $<$4.7 & $<$7.8 & $<$9.5 & $<$11.0 \\
R0142 & 01 42 03.4 & +21 31 15 & 0.280 & L  & $<$13.0 & $<$17.0& $<$14.0 & $<$18.9 & $<$20.4 \\
A0267 & 01 52 41.9 & +01 00 27 & 0.231 & HL  & $<$2.5 & $<$4.8 & $<$8.0 & $<$9.6 & $<$11.2 \\
A0291 & 02 01 43.1 & --02 11 47 & 0.196 & L  & $<$13.0 & $<$17.0 & $<$14.0 & $<$18.9 & $<$20.4 \\
RCS0224 & 02 24 34.2 & --00 02 32 & 0.773 & H  & $<$2.3 & $<$4.7 & $<$8.4 & $<$9.6 & $<$10.9 \\
A0368 & 02 37 27.8 & --26 30 29 & 0.220 & H  & $<$2.5 & $<$4.6 & $<$7.7 & $<$8.9 & $<$10.2 \\
A0383 & 02 48 03.4 & --03 31 44 & 0.187 & HL  & 12.3 $\pm$  0.1 & 32.5 $\pm$  0.5 & 14.9 $\pm$  6.3 & 14.6 $\pm$  6.0 & $<$10.6 \\
A3084 & 03 04 03.9 & --36 56 27 & 0.219 & H  & $<$2.4 & $<$4.4 & $<$7.7 & $<$9.1 & $<$10.7 \\
A3088 & 03 07 02.1 & --28 39 58 & 0.253 & H  &  6.9 $\pm$  0.2 &  6.1 $\pm$  0.5 & $<$7.0 & $<$8.5 & $<$10.1 \\
MACS0451 & 04 51 54.6 & +00 06 18 & 0.430 & H  & $<$2.3 & $<$4.2 & $<$13.0 & $<$15.6 & $<$19.5 \\
A0521 & 04 54 06.9 & --10 13 24 & 0.253 & H  & $<$2.3 & $<$4.5 & $<$9.4 & $<$10.2 & $<$11.9 \\
AS0592 & 06 38 45.2 & --53 58 22 & 0.222 & H  & $<$2.4 & $<$4.7 & $<$15.3 & $<$15.7 & $<$16.3 \\
MACS0647 & 06 47 50.7 & +70 14 54 & 0.591 & H  & $<$2.3 & $<$4.8 & $<$14.7 & $<$14.2 & $<$14.5 \\
BULLET & 06 58 38.1 & --55 57 26 & 0.296 & H  & $<$3.1 & $<$5.8 & $<$8.2 & $<$9.6 & $<$12.4 \\
MACS0717 & 07 17 37.2 & +37 44 23 & 0.546 & H  & $<$2.3 & $<$4.3 & $<$10.5 & $<$11.1 & $<$11.7 \\
A0586 & 07 32 20.3 & +31 38 01 & 0.171 & L  & $<$13.0 & $<$17.0 & $<$14.0 & $<$18.9 & $<$20.4 \\
Z1432 & 07 43 23.1 & +17 33 42 & 0.111 & L  & $<$13.0 & $<$17.0 & -- & -- & -- \\
MACS0744 & 07 44 52.8 & +39 27 25 & 0.698 & H  & $<$2.3 & $<$4.6 & $<$7.9 & $<$9.5 & $<$10.7 \\
A0611 & 08 00 56.8 & +36 03 24 & 0.288 & H  & $<$2.3 & $<$4.5 & $<$8.8 & $<$9.6 & $<$10.7 \\
Z1693 & 08 25 57.8 & +04 14 48 & 0.225 & L  & $<$13.0 & $<$17.0 & $<$14.0 & $<$18.9 & $<$20.4 \\
A0665 & 08 30 57.3 & +65 50 32 & 0.182 & L  & $<$13.0 & $<$17.0 & $<$14.0 & $<$18.9 & $<$20.4 \\
A0689 & 08 37 24.6 & +14 58 22 & 0.279 & L  & -- & -- & $<$14.0 & $<$18.9 & $<$20.4 \\
Z1883 & 08 42 55.9 & +29 27 27 & 0.194 & L  & 15.3 $\pm$  2.2 & 15.4 $\pm$  2.9 & 10.2 $\pm$ 5.2 & $<$18.9 & $<$20.4 \\
A0697 & 08 42 57.6 & +36 22 00 & 0.282 & HL  & $<$2.3 & $<$4.2 & $<$8.2 & $<$10.0 & $<$11.7 \\
Z2089 & 09 00 36.9 & +20 53 40 & 0.235 & L  & 55.5 $\pm$  2.5 & 24.3 $\pm$  2.9 & 12.4 $\pm$  5.2 &  5.8 $\pm$  4.3 & $<$20.4 \\
A0773S & 09 17 53.4 & +51 43 39 & 0.217 & H  & $<$2.4 & $<$4.3 & $<$8.0 & $<$9.6 & $<$11.2 \\
A0773N & 09 17 53.5 & +51 44 01 & 0.217 & H  & $<$2.4 & $<$4.3 & $<$8.0 & $<$9.6 & $<$11.2 \\
A0851 & 09 42 57.5 & +46 58 50 & 0.407 & H  & 23.7 $\pm$  0.3 & 25.3 $\pm$  0.7 & 20.5 $\pm$  5.5 & 12.3 $\pm$  6.2 & $<$11.0 \\
A0868 & 09 45 26.4 & --08 39 07 & 0.153 & H  & $<$2.4 & $<$5.2 & $<$7.7 & $<$8.9 & $<$10.6 \\
Z2701 & 09 52 49.2 & +51 53 06 & 0.214 & H  & $<$2.4 & $<$4.7 & $<$7.1 & $<$8.2 & $<$9.8 \\
A0963 & 10 17 03.6 & +39 02 49 & 0.206 & HL  & $<$2.5 & $<$4.9 & $<$7.9 & $<$10.0 & $<$11.2 \\
MACS1149 & 11 49 35.7 & +22 23 55 & 0.544 & H  & $<$2.4 & $<$4.4 & $<$7.6 & $<$9.0 & $<$10.5 \\
A1413 & 11 55 17.8 & +23 24 16 & 0.143 & H  & $<$2.4 & $<$4.6 & $<$7.7 & $<$8.9 & $<$10.6 \\
A1689 & 13 11 29.5 & --01 20 28 & 0.183 & L  & $<$13.0 & $<$17.0 & $<$14.0 & $<$18.9 & $<$20.4 \\
A1703 & 13 15 05.3 & +51 49 02 & 0.281 & H  & $<$2.4 & $<$4.5 & $<$7.9 & $<$9.4 & $<$11.8 \\
A1758N & 13 32 38.4 & +50 33 35 & 0.279 & HL  & $<$2.4 & $<$4.9 & $<$8.0 & $<$9.4 & $<$11.3 \\
A1758S & 13 32 52.1 & +50 31 34 & 0.279 & HL  & $<$2.4 & $<$4.9 & $<$8.0 & $<$9.4 & $<$11.3 \\
A1763 & 13 35 20.1 & +41 00 04 & 0.228 & L  & $<$13.0 & $<$17.0 & $<$14.0 & $<$18.9 & $<$20.4 \\
A1835 & 14 01 02.1 & +02 52 43 & 0.253 & L  & 315.1 $\pm$  1.8 & 322.6 $\pm$  2.9 & 122.0 $\pm$  5.2 & 45.9 $\pm$  4.3 & 21.5 $\pm$  6.2 \\
MACS1423 & 14 23 47.9 & +24 04 42 & 0.543 & H  & 13.0 $\pm$  0.1 & 27.7 $\pm$  0.3 & 20.8 $\pm$  4.6 & 14.4 $\pm$  5.8 & $<$10.2 \\
A1914 & 14 25 56.6 & +37 48 59 & 0.171 & HL  & $<$2.5 & $<$5.0 & $<$7.9 & $<$9.0 & $<$11.0 \\
Z7160 & 14 57 15.1 & +22 20 35 & 0.258 & L  & $<$13.0 & $<$17.0 & $<$14.0 & $<$18.9 & $<$20.4 \\
A2218 & 16 35 49.1 & +66 12 44 & 0.171 & L  & $<$13.0 & $<$17.0 & $<$14.0 & $<$18.9 & $<$20.4 \\
A2219 & 16 40 19.8 & +46 42 42 & 0.228 & L  & $<$13.0 & $<$17.0 & $<$14.0 & $<$18.9 & $<$20.4 \\
RXJ1720 & 17 20 10.1 & +26 37 32 & 0.164 & HL  &  9.3 $\pm$  0.1 &  7.2 $\pm$  0.9 & $<$13.6 & $<$14.5 & $<$16.2 \\
A2345 & 21 27 13.7 & --12 09 46 & 0.176 & L  & $<$13.0 & $<$17.0 & $<$14.0 & $<$18.9 & $<$20.4 \\
MACS2129 & 21 29 26.2 & --07 41 29 & 0.589 & H  & $<$2.4 & $<$5.3 & $<$29.7 & $<$27.7 & $<$27.2 \\
RXJ2129 & 21 29 40.0 & +00 05 21 & 0.235 & HL  &  3.9 $\pm$  0.1 &  6.5 $\pm$  0.3 &  7.1 $\pm$  1.8 & $<$10.1 & $<$12.0 \\
MS2137 & 21 40 15.1 & --23 39 40 & 0.313 & H  & $<$2.4 & $<$4.6 & $<$8.0 & $<$9.3 & $<$10.3 \\
A2390 & 21 53 36.8 & +17 41 45 & 0.233 & L  & 17.5 $\pm$  2.4 & 38.6 $\pm$  5.7 & 31.4 $\pm$  5.2 & 20.4 $\pm$  4.3 &  8.2 $\pm$  6.2 \\
A2485 & 22 48 31.1 & --16 06 26 & 0.247 & L  & $<$13.0 & $<$17.0 & $<$14.0 & $<$18.9 & $<$20.4 \\
AS1063 & 22 48 44.0 & --44 31 51 & 0.348 & H  & $<$2.3 & $<$4.3 & $<$8.0 & $<$9.9 & $<$11.5 \\
AS1077 & 22 58 48.3 & --34 48 07 & 0.312 & H  & $<$2.4 & $<$4.6 & $<$8.0 & $<$9.2 & $<$10.0 \\
A2537 & 23 08 22.3 & --02 11 33 & 0.295 & H  & $<$2.5 & $<$5.1 & $<$11.9 & $<$12.8 & $<$12.6 \\
RCS2319 & 23 19 53.4 & +00 38 13 & 0.897 & H  & $<$2.3 & $<$4.5 & $<$7.7 & $<$9.0 & $<$10.4 \\
RCS2327 & 23 27 27.6 & --02 04 37 & 0.700 & H  & $<$2.5 & $<$4.8 & $<$9.6 & $<$11.0 & $<$13.5 \\
A2631 & 23 37 39.7 & +00 16 17 & 0.273 & H  & $<$2.4 & $<$4.7 & $<$8.2 & $<$9.5 & $<$10.8 \\
A2667 & 23 51 39.4 & --26 05 03 & 0.230 & H  & 20.4 $\pm$  0.1 & 27.4 $\pm$  0.4 & 10.6 $\pm$  2.6 & $<$9.5 & $<$11.9 \\
\enddata
\tablenotetext{a}{Cluster name, including A/B designation (as in the literature) for multiple candidate BCGs in a single cluster}
\tablenotetext{b}{Position of the BCG in optical (HST) imaging}
\tablenotetext{c}{Subsample: H$=$HLS; L$=$LoCuSS. For clusters in both samples, HL is shown for reference, although the deeper HLS Herschel images have provided the photometry.}
\label{tab:phot}
\end{deluxetable}

\placetable{tab:seds}
\begin{deluxetable}{lcccccc}
\tablecolumns{10}
\tablewidth{0pc}
\tablecaption{Derived SED parameters for Herschel and/or H$\alpha$ detected BCGs.}
\tablehead{\colhead{Cluster ID\tablenotemark{a}} & \colhead{H/L/E\tablenotemark{b}} & \colhead{$L_{\rm TIR}(24)$\tablenotemark{c}} & \colhead{$SFR_{\rm 24}$\tablenotemark{c}} & \colhead{$L_{\rm TIR}(FIR)$\tablenotemark{d}} & \colhead{$SFR_{\rm FIR}$\tablenotemark{d}} & \colhead{$L({\rm H}\alpha)$\tablenotemark{e}} \\ \colhead{} & \colhead{} & \colhead{(L$_\sun$)} & \colhead{(M$_\sun$ yr$^{-1}$)} & \colhead{(L$_\sun$)} & \colhead{(M$_\sun$ yr$^{-1}$)} & \colhead{log(erg s$^{-1}$)}}
\startdata
Z0348 & L & 2.9 $\pm$ 0.07 $\times$10$^{11}$ & 50.5 $\pm$ 1.1 & 1.9 $\pm$ 0.06$\times$10$^{11}$ & 32.6 $\pm$ 1.2 & 42.50 \\
A0383 & HL & 9.4 $\pm$ 0.51 $\times$10$^{9}$ & 1.6 $\pm$ 0.1 & 2.3 $\pm$ 0.10$\times$10$^{10}$ & 4.0 $\pm$ 0.2 & 41.85 \\
A3088 & H & 1.3 $\pm$ 0.03 $\times$10$^{10}$ & 2.2 $\pm$ 0.1 & 1.8 $\pm$ 0.06$\times$10$^{10}$ & 3.0 $\pm$ 0.1 & 41.30 \\
Z1883 & L & $<$1.3$\times$10$^{10}$ & $<$2 & 2.4 $\pm$ 0.32 $\times$10$^{10}$ & 4.1 $\pm$ 0.5 & 42.14 \\
Z2089 & L & 2.0 $\pm$ 0.12 $\times$10$^{12}$ & 346.4 $\pm$ 20.6 & 1.2 $\pm$ 0.06 $\times$10$^{11}$ & 20.2 $\pm$ 1.0 & 42.50 \\
A0851 & H & 5.7 $\pm$ 0.19 $\times$10$^{10}$ & 9.9 $\pm$ 0.3 & 1.9 $\pm$ 0.09$\times$10$^{11}$ & 32.7 $\pm$ 1.6 & -- \\
Z3146 & E & 3.5 $\pm$ 0.34 $\times$10$^{11}$ & 61.0 $\pm$ 6.0 & 5.4 $\pm$ 0.29 $\times$10$^{11}$ & 93.1 $\pm$ 5.1 & 42.55 \\
A1068 & E & 1.5 $\pm$ 0.04 $\times$10$^{12}$ & 254.4 $\pm$ 6.8 & 5.8 $\pm$ 0.03 $\times$10$^{11}$ & 99.3 $\pm$ 0.5 & 41.94 \\
A1835 & L & 1.2 $\pm$ 0.01 $\times$10$^{12}$ & 202.0 $\pm$ 0.5 & 8.5 $\pm$ 0.05 $\times$10$^{11}$ & 146.5 $\pm$ 0.8 & 42.20 \\
MACS1423 & H & 1.4 $\pm$ 0.04 $\times$10$^{11}$ & 24.8 $\pm$ 0.7 & 2.7 $\pm$ 0.17 $\times$10$^{11}$ & 46.0 $\pm$ 2.9 & 42.00 \\
RXJ1720 & HL & 1.2 $\pm$ 0.19 $\times$10$^{10}$ & 2.0 $\pm$ 0.3 & 9.3 $\pm$ 0.27 $\times$10$^{9}$ & 1.6 $\pm$ 0.1 & 40.88 \\
RXJ2129 & HL & 6.0 $\pm$ 0.25 $\times$10$^{9}$ & 1.0 $\pm$ 0.1 & 1.0 $\pm$ 0.05 $\times$10$^{10}$ & 1.7 $\pm$ 0.1 & 40.75 \\
A2390 & L & 5.1 $\pm$ 0.03 $\times$10$^{10}$ & 8.8 $\pm$ 0.1 & 4.9 $\pm$ 0.49 $\times$10$^{10}$ & 8.5 $\pm$ 0.7 & 41.69 \\
A2597 & E & 1.2 $\pm$ 0.12 $\times$10$^{10}$ & 2.1 $\pm$ 0.2 & 1.8 $\pm$ 0.05 $\times$10$^{10}$ & 3.1 $\pm$ 0.1 & 41.48 \\
A2667 & H & 1.8 $\pm$ 0.02 $\times$10$^{11}$ & 31.2 $\pm$ 0.3 & 5.0 $\pm$ 0.14 $\times$10$^{10}$ & 8.7 $\pm$ 0.2 & 41.60 \\
\hline
\hline
A0115 & L & 1.8 $\pm$ 0.04 $\times$10$^{10}$ & 3.0 $\pm$ 0.1 & $<$2.5$\times$10$^{10}$ & $<$4 & 41.03 \\
A0291 & L & 1.8 $\pm$ 0.04 $\times$10$^{10}$ & 3.2 $\pm$ 0.1 & $<$2.5$\times$10$^{10}$ & $<$4 & 41.36 \\
Z2701 & H & $<$4.2$\times$10$^{9}$ & $<$1 & $<$6.4$\times$10$^{9}$ & $<$1 & 40.64 \\
Z7160 & L & 2.4 $\pm$ 0.04 $\times$10$^{10}$ & 4.2 $\pm$ 0.1 & $<$4.2$\times$10$^{10}$ & $<$9.6 & 41.40 \\
\enddata
\tablenotetext{a}{Cluster ID as in Table \ref{tab:phot}}
\tablenotetext{b}{Subsample: H$=$HLS; L$=$LoCuSS; E$=$\citet{edg10-47}. Clusters in both HLS and LoCuSS are indicated, although the deeper HLS Herschel provided the photometry.}
\tablenotetext{c}{Estimated from the 24 \micron{} photometry (via \citealt{rie09-556})}
\tablenotetext{d}{Derived from the best fit \citet{rie09-556} template}
\tablenotetext{e}{{}Literature H$\alpha$ luminosity}
\\
BCGs listed beneath the dividing line were not detected in any of the FIR Herschel bands, but do have a measured H$\alpha$ emission line flux
\label{tab:seds}
\end{deluxetable}

\appendix
\section{Infrared SED fits}
\label{app:seds}

\begin{figure*}
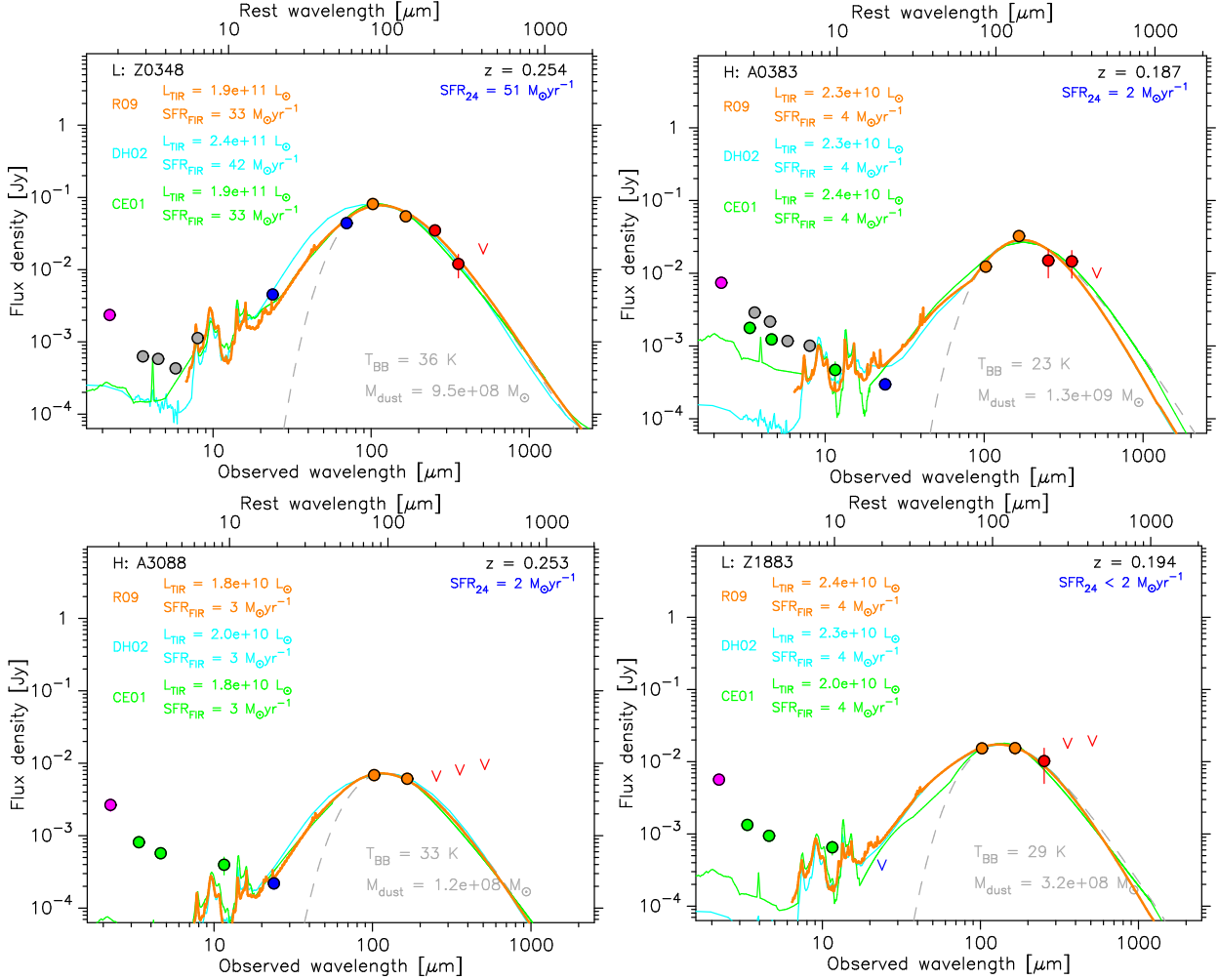

\includegraphics[angle=270,scale=0.65]{fig9_a.eps}
\includegraphics[angle=270,scale=0.65]{fig9_b.eps}
\includegraphics[angle=270,scale=0.65]{fig9_c.eps}
\includegraphics[angle=270,scale=0.65]{fig9_d.eps}
\caption{Far-infrared SEDs for Herschel-detected BCGs (sample, name and redshift shown in black). PACS and SPIRE detections displayed as orange and red circles respectively. Upper limits are shown as `v' at the 3$\,\sigma$ detection limit. Additional photometry is also displayed: 2MASS $K$-band (magenta); {\it Spitzer} IRAC (grey) and MIPS (blue); WISE (green); IRAS (cyan); SCUBA (black). A modified-blackbody is fit to the Herschel points (dashed line), with the temperature and estimated dust mass shown in the lower-right corner. Best fit templates from three libraries are also plotted: orange $=$ \citet{rie09-556}, cyan $=$ \citet{dal02-159} and green $=$ \citet{cha01-562}. Derived $L_{\rm TIR}$ and SFR$_{\rm FIR}$ for each are given in matching colors. MIPS 24 \micron{}-derived SFR is shown in blue in the upper-right corner.}
\label{fig:seds1}
\end{figure*}

\begin{figure*}
\includegraphics[angle=270,scale=0.65]{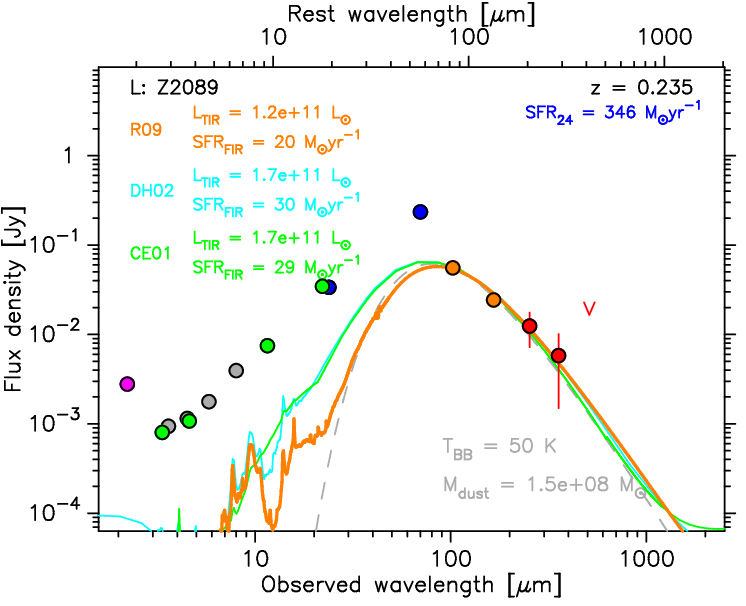}
\includegraphics[angle=270,scale=0.65]{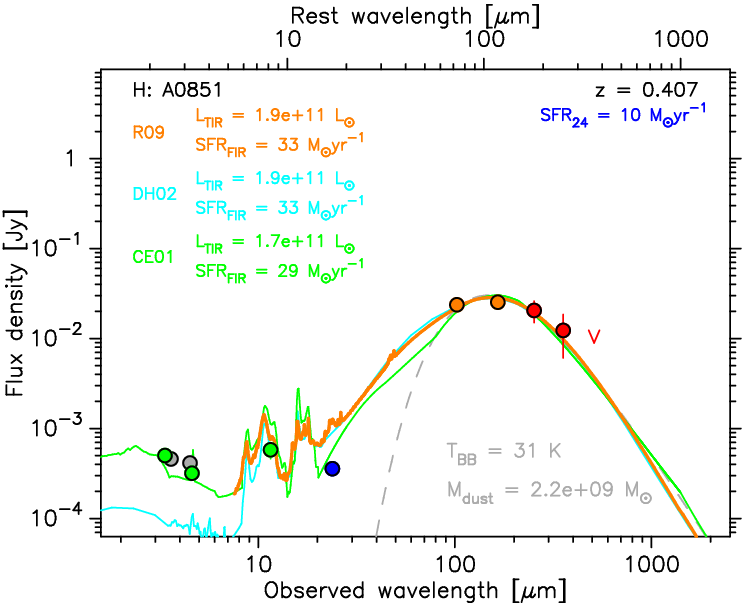}
\includegraphics[angle=270,scale=0.65]{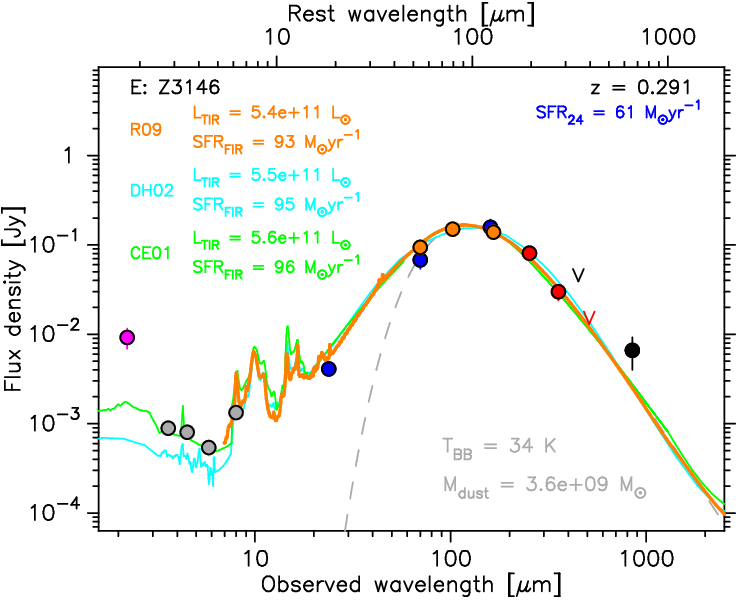}
\includegraphics[angle=270,scale=0.65]{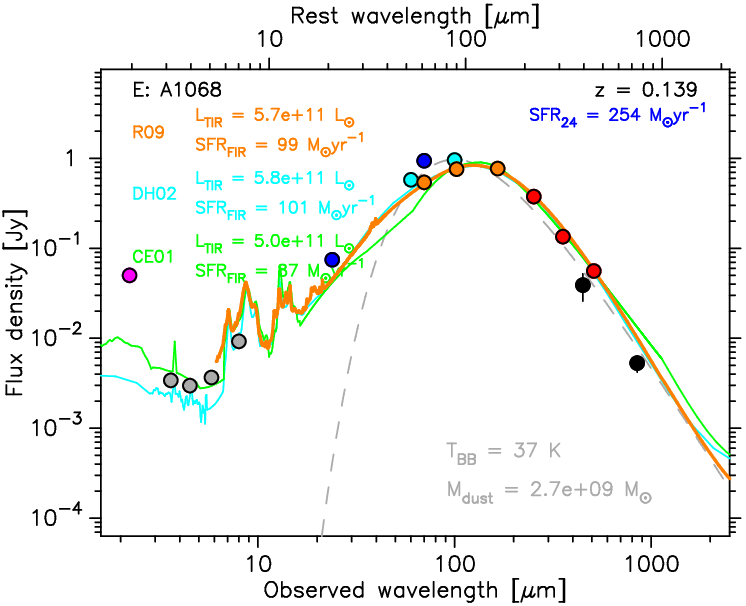}
\includegraphics[angle=270,scale=0.65]{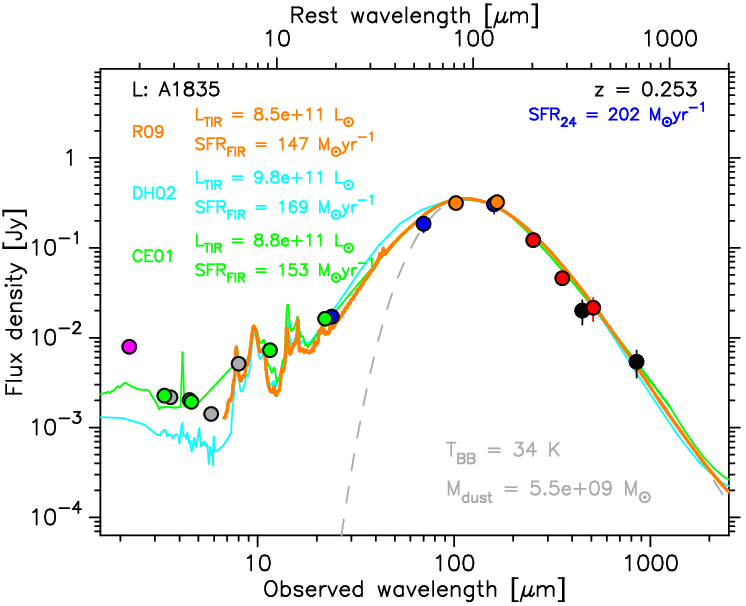}
\includegraphics[angle=270,scale=0.65]{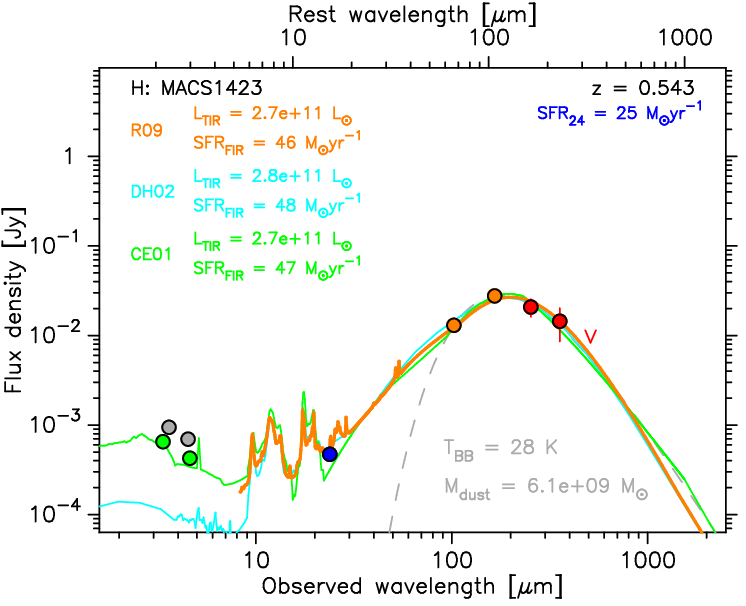}
\caption{Far-infrared SEDs for Herschel detected BCGs. Layout as in Figure \ref{fig:seds1}.}
\label{fig:seds2}
\end{figure*}

\begin{figure*}
\includegraphics[angle=270,scale=0.65]{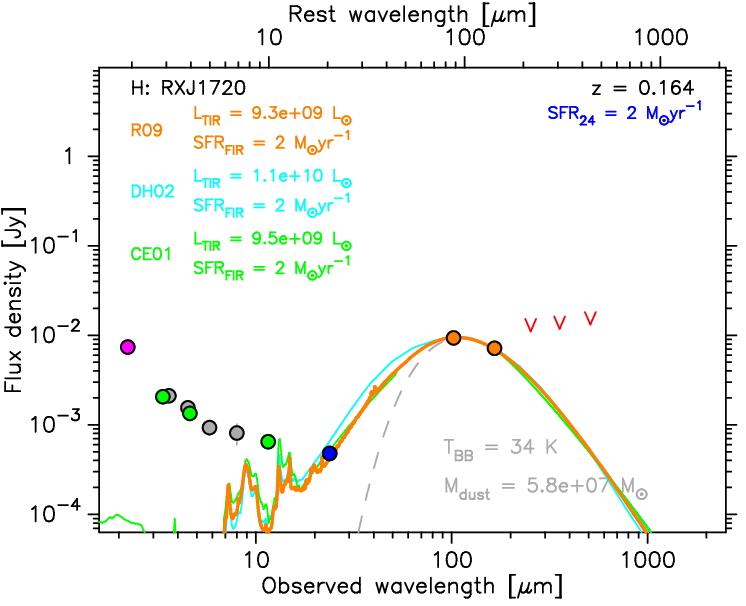}
\includegraphics[angle=270,scale=0.65]{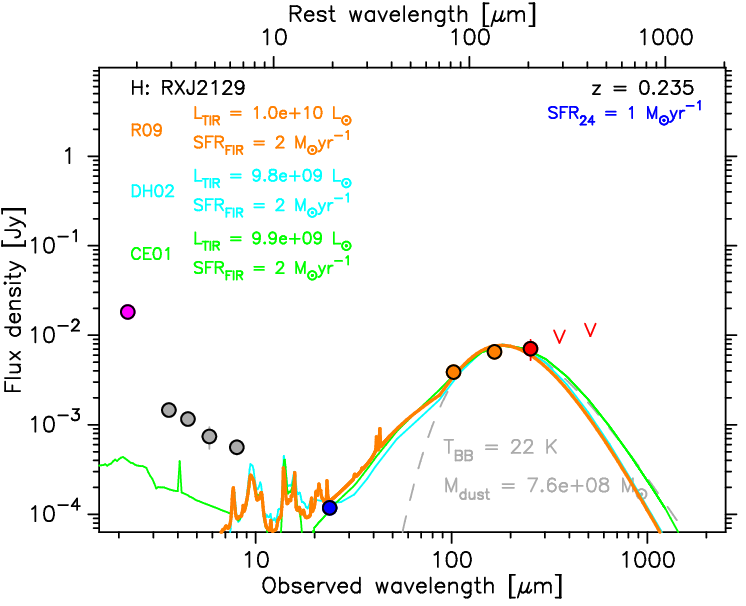}
\includegraphics[angle=270,scale=0.65]{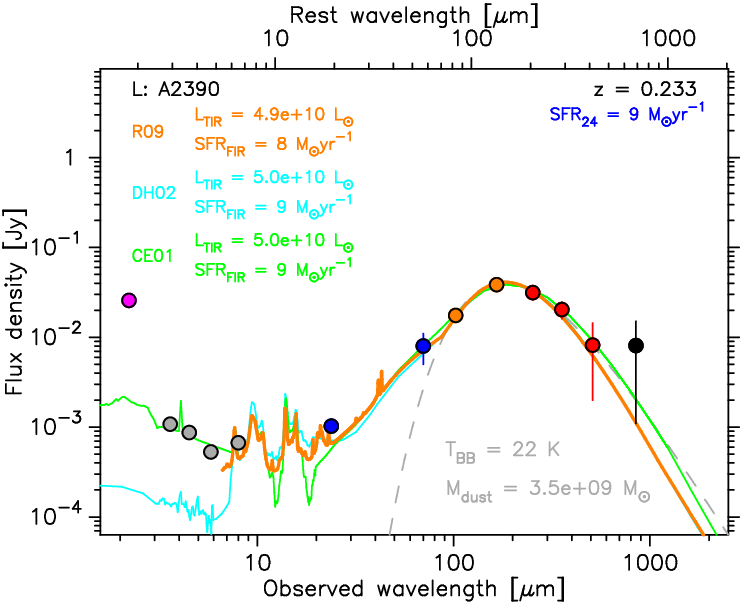}
\includegraphics[angle=270,scale=0.65]{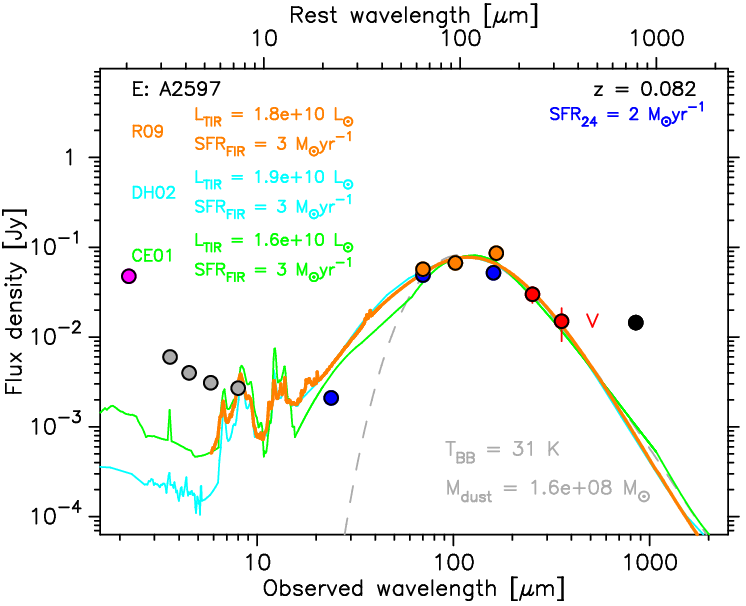}
\includegraphics[angle=270,scale=0.65]{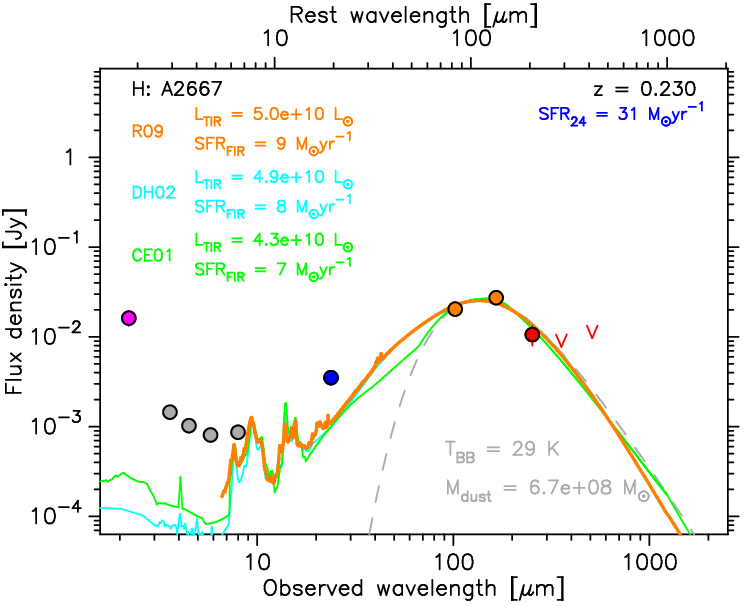}
\caption{Far-infrared SEDs for Herschel detected BCGs. Layout as in Figure \ref{fig:seds1}.}
\label{fig:seds3}
\end{figure*}

Figures \ref{fig:seds1}--\ref{fig:seds3} present the infrared (observed frame $\lambda$ $=$ 2--2000 \micron) spectral energy distribution for the 15 Herschel-detected BCGs in HLS, LoCuSS and E10.

\placefigure{fig:seds}

\end{document}